# Filtering Random Graph Processes Over Random Time-Varying Graphs

Elvin Isufi, Andreas Loukas, Andrea Simonetto, and Geert Leus

*Abstract*—**Graph filters play a key role in processing the graph spectra of signals supported on the vertices of a graph. However, despite their widespread use, graph filters have been analyzed only in the deterministic setting, ignoring the impact of stochasticity in both the graph topology as well as the signal itself. To bridge this gap, we examine the statistical behavior of the two key filter types, finite impulse response (FIR) and autoregressive moving average (ARMA) graph filters, when operating on random time-varying graph signals (or random graph processes) over random time-varying graphs. Our analysis shows that (*i*) in expectation, the filters behave as the same deterministic filters operating on a deterministic graph, being the expected graph, having as input signal a deterministic signal, being the expected signal, and (*ii*) there are meaningful upper bounds for the variance of the filter output. We conclude the paper by proposing two novel ways of exploiting randomness to improve (joint graph-time) noise cancellation, as well as to reduce the computational complexity of graph filtering. As demonstrated by numerical results, these methods outperform the disjoint average and denoise algorithm, and yield a (up to) four times complexity redution, with very little difference from the optimal solution.**

## I. Introduction

As witnessed by their widespread use [1], graphs are proving to be one of the most successful models for capturing the complex relations featured for example by social networks [2], sensor networks [3], and biological networks [4]. Recently, in the signal processing community, an important effort has been put to extend standard signal processing concepts from classical domains, like time and space (i.e., for time series and images), to signals supported on the vertices of an irregular graph. This extension has important repercussions in many of the aforementioned fields, and it is generally known as signal processing on graphs [5].

The main tool of signal processing on graphs is the graph Fourier transform [6] (GFT). By establishing a notion of frequency appropriate for graphs, the GFT enables the analysis of graph signals in the graph frequency domain, rather than only in the vertex domain. Together with the graph Fourier transform, *graph filters* arise as an inseparable duet. Graph filters are the linear operators that amplify or attenuate specific parts of a graph signal's frequency spectrum. Distinguished applications of graph filters are, for instance, customer behavior prediction [6], signal compression and data classification [6], as well as signal smoothing, denoising and interpolation [7]–[9]. To date, two main graph filter types are known: *finite* (FIR) and *infinite* (IIR) *impulse response* graph filters. FIR graph filters provide sufficient approximation accuracy and have a simple implementation, but may suffer in time-varying scenarios, e.g., in cases of time-varying signals and/or time-varying graphs. On the other hand, at the expense of an increased computational complexity, IIR graph filters [10]–[13] can handle better time-variations and they give exact solutions to some specific denoising [8] and interpolation [9] problems.

Despite their popularity, graph filter analysis has so far been exclusively deterministic. Albeit, in applications like communication networks, social networks, smart grids, and road networks, stochasticity is of paramount importance: it occurs for instance when the graph signal is corrupted with random additive noise, when it obeys a certain distribution, or when –owing to random link and node failures– the graph topology becomes random. Characterizing the impact of random graph perturbations has been considered for example in distributed optimization [14]–[17], but not in the context of graph filters.

This work approaches the analysis of graph filters from a statistical standpoint. Specifically, we analyze the graph filters' behavior when the input signal on the graph *and* the graph topology are random processes over *time* with given statistical properties, yet with independent realizations. In this paper, when we talk about a random graph topology, we mean the signal graph, i.e., the graph that explains the signal structure, which also corresponds to the graph that is used for processing the graph signals. For instance, in a road network, the signal graph can be random due to accidents. The actual graph used for data exchange, named the communication graph (e.g., between sensor nodes placed on the cross roads of a road network) could be similar to the signal graph or not, but we will not discuss this issue in this paper. Hence, in the sequel, the term graph will always refer to the *signal graph*. We believe that computing the filter output by incorporating the stochastic graph realizations is more meaningful, since they are integral part of the actual graph signal realizations. In the road network for instance, the traffic at a particular time instant is a consequence of the past realizations of the graph. Finally, in this work we also view stochasticity as a tool to alleviate the computational and communication costs in distributed graph filtering over a deterministic graph.

We provide the following contributions:

*(i) Graph filters in the mean (Section III). (i-a)* For a random time-varying graph signal characterized by *temporal*

E. Isufi and G. Leus are with the faculty of Electrical Engineering, Mathematics and Computer Science, Delft University of Technology, 2826 CD Delft, The Netherlands. Andreas Loukas is with the faculty of Electrical Engineering of the Swiss Federal Institute of Technology in Lausanne (EPFL), Switzerland. Andrea Simonetto is with the optimisation and control group of IBM Research Ireland. E-mails: {e.isufi-1, g.j.t.leus}@tudelft.nl, andreas.loukas@epfl.ch, andrea.simonetto@ibm.com.

*stationarity with independent realizations*, living on a random time-varying graph, we show that the expected filter output is equal to the output of the same filter operating on the expected graph signal living on the expected graph. *(i-b)* We show that in case of a random time-varying, yet *non-stationary* graph signal with independent realizations, (i.e., a time-varying mean and covariance) over a random time-varying graph, the mean of the output signal is equal to the output of a deterministic 2-dimensional filter, operating on the time-varying mean signal over the mean graph. In the latter case, the filter jointly captures the variations in the mean of the graph signal over the graph and time domains.

*(ii) Variance analysis (Section IV).* We show that the average node variance of the filter output is upper bounded. This generalizes the known result for a first-order autoregressive moving average (ARMA) graph filter [18]. We also propose a recursive way to track the variance of the ARMA output iteratively based only on statistical knowledge of the previous time instant. Our expressions for the first and second order statistics of the filter output, demonstrate that state-of-the-art graph filters are equipped to handle stochastic settings (for graphs and signals). Nevertheless, graph filters of high order should be used with caution, since the output signal variance increases with the filter order.

*(iii) Leveraging stochasticity (Section V).* (iii-a) We propose to use stochasticity to improve graph signal denoising tasks under smoothness priors, a.k.a. Tikhonov denoising, when multiple realizations of the graph signal are available. The proposed approach performs an online joint denoising over the graph and time domains exploiting the new realizations of the input signal. (iii-b) We also make use of stochasticity to perform classical graph filtering with lower communication and complexity costs. The proposed approach considers performing the filtering over a sparsified graph where the information is exchanged only with some randomly chosen neighbors. This can be seen as an intersection between sparsification [19] and gossiping [20]. As our bounds and simulations suggest, such an approach performs better with low-order graph filters and has the potential to reduce the communication and computational costs up to 75%.

Our analysis is validated by numerical simulations in Section VI. Finally, the paper is concluded in Section VII.

The proofs are presented in Appendix.

**Notation.** We indicate scalars by normal letters (i.e., $a$ or $A$); vector variables with bold lowercase letters $\boldsymbol{a}$ and matrix variables with bold uppercase letters $\boldsymbol{A}$. Scalars $a_i$ and $A_{ij}$ correspond to the entries of $\boldsymbol{a}$ and $\boldsymbol{A}$, respectively. We indicate by $|a|$ the absolute value of $a$ and by $\|\boldsymbol{a}\|$ and $\|\boldsymbol{A}\|$ the 2-norm and the spectral norm of the vector $\boldsymbol{a}$ and matrix $\boldsymbol{A}$, respectively. To characterize convergence, we adopt the term *linear convergence*, which asserts that a recursion converges to its stationary value exponentially with time (i.e., linearly in a logarithmic scale) [21]. To ease the notation, we indicate with $\mathbb{E}[\cdot]$, $cov(\cdot)$ and $var(\cdot)$ the expectation, the covariance and variance operator, respectively and, moreover, $\boldsymbol{\Sigma}_{\boldsymbol{a}}[t] = \mathbb{E}\big[\boldsymbol{a}_t\boldsymbol{a}_t^{\mathsf{H}}\big] - \mathbb{E}[\boldsymbol{a}_t]\,\mathbb{E}[\boldsymbol{a}_t]^{\mathsf{H}}$ refers to the covariance matrix at time instant $t$ of the random process $\boldsymbol{a}_t$. The trace operator of a matrix is indicated as $tr(\cdot)$.

## II. Preliminaries

In this Section, we recap the basics of the graph Fourier transform and graph filters. Then, we explain the stochastic model used for the graph topology and graph signal.

### A. Background

Consider an undirected graph $\mathcal{G} = (\mathcal{V}, \mathcal{E})$ of $N$ nodes and $M$ edges, where $\mathcal{V}$ indicates the set of nodes (vertices) and $\mathcal{E}$ the set of edges. Let $\boldsymbol{x}$ be the graph signal supported on set $\mathcal{V}$, whose $i$-th component $x_i \in \mathbb{C}$ represents the value of the signal at the $i$-th node, denoted as $u_i \in \mathcal{V}$.

**Graph Fourier transform (GFT).** The GFT transforms a graph signal $\boldsymbol{x}$ into the graph frequency domain $\hat{\boldsymbol{x}}$: the forward and inverse GFTs are $\hat{\boldsymbol{x}} = \boldsymbol{\Phi}^\top \boldsymbol{x}$ and $\boldsymbol{x} = \boldsymbol{\Phi}\hat{\boldsymbol{x}}$, where the columns of the unitary matrix $\boldsymbol{\Phi}$, indicated as $\boldsymbol{\phi}_n$, form an orthonormal basis and are commonly chosen as the eigenvectors of a graph Laplacian $\boldsymbol{L}$, such as the discrete Laplacian $\boldsymbol{L}_{\mathrm{d}}$ or normalized Laplacian $\boldsymbol{L}_{\mathrm{n}}$. The corresponding eigenvalues are denoted as $\{\lambda_n\}_{n=1}^N$ and will indicate the graph frequencies. For an extensive review on the properties of the GFT, we refer to [5], [6]. To avoid any restrictions on the generality of our approach, in the following we present our results for a *general Laplacian matrix* $\boldsymbol{L}$, which is *symmetric* and *local*: for all $i \neq j$, $L_{ij} = 0$ whenever $u_i$ and $u_j$ are not neighbors and $L_{ij} = L_{ji}$ otherwise[1]. We derive our results for a class of graphs with general Laplacian matrices in some restricted set $\mathcal{L}$. We assume that for every $\boldsymbol{L} \in \mathcal{L}$ the minimum eigenvalue is bounded below by $\lambda_{min}$ and the maximum eigenvalue is bounded above by $\lambda_{max}$. Hence, all considered graphs have a bounded spectral norm, i.e., $\|\boldsymbol{L}\| \leq \varrho = \max\{|\lambda_{min}|, |\lambda_{max}|\}$. For instance, when $\boldsymbol{L} = \boldsymbol{L}_{\mathrm{d}}$, we can take $\lambda_{min} = 0$ and $\lambda_{max}$ related to the maximum eigenvalue of any of the graphs. When $\boldsymbol{L} = \boldsymbol{L}_{\mathrm{n}}$, we can consider $\lambda_{min} = 0$ and $\lambda_{max} = 2$. As in [13], also translated Laplacians such as $\boldsymbol{L} = \boldsymbol{L}_{\mathrm{d}} - \lambda_{max}/2\boldsymbol{I}$ or $\boldsymbol{L} = \boldsymbol{L}_{\mathrm{n}} - \boldsymbol{I}$ can be considered.

**Graph filters.** A *graph filter* $\boldsymbol{H}$ is an operator that acts upon a graph signal $\boldsymbol{x}$ by amplifying or attenuating its graph Fourier coefficients as

$$\boldsymbol{H}\boldsymbol{x} = \sum_{n=1}^{N} H(\lambda_n)\, \hat{x}_n\, \boldsymbol{\phi}_n. \tag{1}$$

The graph frequency response $H : [\lambda_{min}, \lambda_{max}] \to \mathbb{C}$ controls how much $\boldsymbol{H}$ amplifies the signal graph spectra $H(\lambda_n) = (\boldsymbol{\phi}_n^\top \boldsymbol{H}\boldsymbol{x})/\hat{x}_n$. We are interested in how we can filter a signal with a graph filter $\boldsymbol{H}$ having a user-provided frequency response $H^*(\lambda)$. In the following two paragraphs we recall two ways to do so, namely the FIR and ARMA graph filters, where the latter belongs to the class of IIR graph filters [12], [13]. Since we will often consider random time-varying graph topologies in this work, without any knowledge on the eigenvalue spread for different time instants, we will focus on universal graph filters, i.e., where the filter design is performed for a continuous range of graph frequencies $\lambda$ like the FIR design in [7] or the ARMA design in [13]. However, in case the eigendecomposition cost is affordable and the eigenvalues of all graph realizations are known, it is recommended to perform the filter design for that particular set of graph frequencies [6], [23], [24] instead of adopting the universal approach. This will of course lead to an enhanced approximation accuracy of the filter.

[1]Note that even though we limit ourselves to undirected graphs, the core idea can be extended to directed graphs using the adjacency matrix [22].

**FIR graph filters.** A universal FIR graph filter of order $K$ implements the filtering operation by approximating a graph filter $\boldsymbol{H}$ using a $K$-th order polynomial of $\boldsymbol{L}$ [7]. The output signal of a $K$-th order FIR filter (denoted as $\text{FIR}_K$) and the graph frequency response are respectively given by

$$\boldsymbol{z} = \sum_{k=0}^{K} \varphi_k \boldsymbol{L}^k \boldsymbol{x} \quad \text{and} \quad H(\lambda) = \sum_{k=0}^{K} \varphi_k \lambda^k. \tag{2}$$

The coefficients $\varphi_k$ can be, for instance, found by a Chebyshev polynomial approximation or least-squares fitting to the desired frequency response $H^*(\lambda)$ [7]. The computation of $\text{FIR}_K$ is easily performed distributedly. Since $\boldsymbol{L}^K\boldsymbol{x} = \boldsymbol{L}\left(\boldsymbol{L}^{K-1}\boldsymbol{x}\right)$, each node $u_i$ can compute the $K$-th term from the values of the $(K-1)$-th term in its neighborhood. The overall communication and computational complexity to perform the filtering distributively is $O(MK)$. We refer the reader to [7] and [13] for further details on the distributed implementation of graph filters.

**ARMA graph filters.** In [12], we introduced ARMA graph filters as a special class of IIR graph filters. Differently from the other IIR graph filters [11], [25], ARMA filters capture naturally time-variations of the graph topology and graph signal, and have well-understood behavior for deterministic time-variations [13] showing larger robustness than FIR filters. The most practical implementation (due to its improved stability) of an ARMA filter of order $K$ ($\text{ARMA}_K$) can be obtained by using a *parallel* bank of $K$ $\text{ARMA}_1$ filters. Let us denote with the superscript $(k)$ the terms that correspond to the $k$-th $\text{ARMA}_1$ filter for $k = 1, 2, \ldots, K$. Then, given a graph signal $\boldsymbol{x}$, the filter memory $\boldsymbol{y}_t$ and output $\boldsymbol{z}_t$ at time instant $t$ are

$$\boldsymbol{y}_{t+1}^{(k)} = \psi^{(k)} \boldsymbol{L} \boldsymbol{y}_t^{(k)} + \varphi^{(k)} \boldsymbol{x} \tag{3a}$$

$$\boldsymbol{z}_{t+1} = \sum_{k=1}^{K} \boldsymbol{y}_{t+1}^{(k)}, \tag{3b}$$

for arbitrary $\boldsymbol{y}_0^{(k)}$, and where $\psi^{(k)}$ and $\varphi^{(k)}$ are the complex filter coefficients. From [13], the graph frequency response of a parallel $\text{ARMA}_K$ filter is

$$H(\lambda) = \sum_{k=1}^{K} \frac{r_k}{\lambda - p_k} \quad \text{subject to} \quad |p_k| > \varrho, \tag{4}$$

where $r_k = -\varphi^{(k)}/\psi^{(k)}$ represents the filter residuals and $p_k = 1/\psi^{(k)}$ the poles. Note that in contrast to FIR filters, where the filter output is obtained after $K$ time instants, recursion (3) will achieve the frequency response (4) theoretically at infinity, yet characterized by a linear convergence in practice. The $\text{ARMA}_K$ has a per-iteration communication and computational complexity of $O(MK)$ [13].

### *B. Stochastic model*

We proceed to formalize the stochastic model used in the rest of this paper. Our definitions here define the considered stochasticity of the graph, as well as the graph signal.

**Graph model.** We consider a random time-varying graph $\mathcal{G}_t$ as a random edge sampling of an arbitrary, but time-invariant underlying graph $\mathcal{G}$. More formally:

*Random edge sampling (RES) graph model.* The probability that a link $(i, j)$ in the edge set $\mathcal{E}$ is activated at time $t$ is $p_{i,j}$, with $0 < p_{i,j} \leq 1$. The edges are activated independently across time. Graph realizations are considered mutually independent with the random graph process.

Thus, at each time step $t$, we draw a graph realization $\mathcal{G}_t = (\mathcal{V}, \mathcal{E}_t)$ from the underlying graph $\mathcal{G} = (\mathcal{V}, \mathcal{E})$, where the edge set $\mathcal{E}_t \subseteq \mathcal{E}$ is generated via an i.i.d. Bernoulli process; this is a standard way of studying link failures in the literature on network algorithms [26], [27]. In a graph signal processing perspective, the RES model suits better cases where the graph has physical meaning, e.g., smart grids, communication networks, or street networks. In these situations the RES model may address grid problems, link failures, or street closures, which have stochastic repercussions on the filtering output. Further, as we will show in Section V-B, this random graph model also results to be a useful tool to perform sparsified graph filtering over a deterministic graph.

From now on, let us refer to $\boldsymbol{L}$ as the Laplacian relative to the underlying graph $\mathcal{G} = (\mathcal{V}, \mathcal{E})$, from which the graph realizations at different time instants are drawn, and to $\boldsymbol{L}_t$ as the Laplacian of the graph realization at time instant $t$. Note that each node locally derives the application of the instantaneous Laplacian matrix $\boldsymbol{L}_t$ on the graph signal $\boldsymbol{x}$ by communicating with its neighbors. For convenience, denote the expected Laplacian $\mathbb{E}[\boldsymbol{L}_t]$ as $\bar{\boldsymbol{L}}$ related to the expected graph $\bar{\mathcal{G}}$.

Since matrices $\boldsymbol{L} \in \mathcal{L}$ and $\mathcal{E}_t \subseteq \mathcal{E}$, the instantaneous graph Laplacian $\boldsymbol{L}_t$ of $\mathcal{G}_t$ belongs also to $\mathcal{L}$, meaning that all $\boldsymbol{L}_t$ have bounded eigenvalues. Further, due to the interlacing property [28], the spectral radius bound $\varrho$ satisfies the property $\|\boldsymbol{L}_t\| \leq \|\boldsymbol{L}\| \leq \varrho$ for all $t$.

**Signal model.** We consider random time-varying graph signals which can be correlated among the nodes for a fixed time instant, but have independent realizations at different time instants. More formally:

*Random graph process model.* The graph signal $\boldsymbol{x}_t \in \mathbb{C}^N$, at time instant $t$, is a realization of a random process, with time-invariant mean $\bar{\boldsymbol{x}}$ and covariance matrix $\boldsymbol{\Sigma}_{\boldsymbol{x}}$. Signal $\boldsymbol{x}_t$ has independent realizations over time.

The random graph process model presented above is a subclass of wide-sense stationary (for short stationary) temporal signals, while not being restricted to stationary graph signals. The graph stationary model has been considered in [29]–[31], where the stationarity of the graph signal $\boldsymbol{x}$ is analyzed only over the graph. Considering the graph Laplacian as a shift, such a signal is generally characterized by a constant mean, i.e., $\bar{\boldsymbol{x}} = \bar{x}\mathbf{1}$, and a covariance matrix $\boldsymbol{\Sigma}_{\boldsymbol{x}}$ that is jointly diagonalizable with the graph Laplacian. In this paper, we work with a broader model, where we do not require the expected graph signal $\bar{\boldsymbol{x}}$ to be a constant vector, or the covariance matrix $\boldsymbol{\Sigma}_{\boldsymbol{x}}$ to be jointly diagonalizable with the graph Laplacian. Further, we would also like to point out the difference with [32], where the authors assume that the covariance matrix of the graph process is related to the underlying graph Laplacian $\boldsymbol{L}$. This approach can be seen as a particular case of the stochastic model presented in this paper, assuming that the covariance matrix of the graph signal is only related to the main graph $\mathcal{G}$ and not to its instantaneous realizations $\mathcal{G}_t$.

In addition to the assumed random graph process model, in Section III-B we extend some of our results to the case where only the second-order moment of the signal is time-invariant, whereas the mean is time-varying. This model encompasses time-varying deterministic signals corrupted with possibly node-correlated noise, such as sensor noise.

## III. Graph filters in the mean

In this section, we analyze the expected behavior of graph filters when the graph topology and graph signal are of stochastic nature. We start by considering our random graph process model and then extend our results to the more general case when the random graph process has a time-varying mean, yet a time-invariant covariance matrix.

### A. Random graph processes

We start with the simpler FIR graph filters. The more involved case of ARMA graph filters is discussed next.

**FIR.** Defining the transition matrix $\boldsymbol{\Phi}_{\boldsymbol{L}}(t', t) := \prod_{\tau=t}^{t'} \boldsymbol{L}_\tau$ if $t' \geq t$ and $\boldsymbol{\Phi}_{\boldsymbol{L}}(t', t) := \boldsymbol{I}$ if $t' < t$, the output $\boldsymbol{z}_{t+1}$ of a $K$-th order FIR graph filter is

$$\boldsymbol{z}_{t+1} = \sum_{k=0}^{K} \varphi_k \boldsymbol{\Phi}_{\boldsymbol{L}}(t, t-k+1)\boldsymbol{x}_{t-k+1}. \tag{5}$$

Notice that, in contrast to the time-invariant setting, the output which is computed at time $t+1$ is a function of all graph realizations in the time interval $[t-K+1, t]$ and graph signals starting from time $t-K+1$ up to time $t+1$. According to the following Proposition, the above graph filter is well behaved when examined in expectation.

*Proposition 1. Consider the FIR$_K$ graph filter (5), the RES graph model and the random graph process model. Then, the expected output of the graph filter after $K$ time steps is*

$$\bar{\boldsymbol{z}}_{t+1} = \sum_{k=0}^{K} \varphi_k \bar{\boldsymbol{L}}^k \bar{\boldsymbol{x}}. \tag{6}$$

Proposition 1 confirms that, in the mean, an FIR$_K$ filter has the same behavior as if this filter was applied to the expected graph, having the expected graph signal as input.

**ARMA.** Let us now move on to ARMA graph filters. The output $\boldsymbol{z}_t$ of the ARMA$_K$ filter at time instant $t$ operating on a time-varying graph signal $\boldsymbol{x}_t$ over a time-varying graph $\mathcal{G}_t$ can be expressed as

$$\boldsymbol{y}_{t+1}^{(k)} = \psi^{(k)} \boldsymbol{L}_t \boldsymbol{y}_t^{(k)} + \varphi^{(k)} \boldsymbol{x}_t \tag{7a}$$

$$\boldsymbol{z}_{t+1} = \sum_{k=1}^{K} \boldsymbol{y}_{t+1}^{(k)}, \tag{7b}$$

where $\boldsymbol{y}_0^{(k)}$ is arbitrary. Assuming time-varying stochasticity, the subscript $t$ indicates the random time-variations of the graph (captured by the Laplacian $\boldsymbol{L}_t$) and input signal $\boldsymbol{x}_t$. Theorem 1 describes the expected behavior of an ARMA$_K$ graph filter.

*Theorem 1. Under the RES graph model, the random graph process model and the stability of the parallel ARMA$_K$ filter, i.e., $|\psi^{(k)}| < 1/\varrho$ for all $k = 1, \ldots, K$, the steady state of the expected value of the ARMA$_K$ recursion (7) is*

$$\bar{\boldsymbol{z}} = \lim_{t \to \infty} \mathbb{E}[\boldsymbol{z}_{t+1}] = \sum_{k=1}^{K} \varphi^{(k)} \left(\boldsymbol{I} - \psi^{(k)} \bar{\boldsymbol{L}}\right)^{-1} \bar{\boldsymbol{x}}. \tag{8}$$

*Recursion (7) converges in the mean to (8) linearly, irrespective of the initial condition $\boldsymbol{y}_0^{(k)}$ and graph realizations $\boldsymbol{L}_t$.*

We can see that (8) is a parallel ARMA$_K$ filter having as input $\bar{\boldsymbol{x}}$ over the graph $\bar{\mathcal{G}}$. Theorem 1 asserts that the expected value of the steady state of the filter output is only influenced by the expected value of the graph signal and by the expected graph, but is independent of any changes in the graph topology.

### B. Random graph processes with time-varying statistics

We can also provide closed-form expressions for the expected filter output in a more general context, i.e., when $\boldsymbol{x}_t$ is characterized by a time-varying mean and covariance. The generalized signal model is:

*Random graph process with time-varying statistics.* The input signal $\boldsymbol{x}_t \in \mathbb{C}^N$, at time instant $t$, is a realization of a random process, with time-dependent first order moment $\bar{\boldsymbol{x}}_t$ and time-dependent covariance matrix $\boldsymbol{\Sigma}_{\boldsymbol{x}}[t]$. Signal $\boldsymbol{x}_t$ has independent realizations over time.

To extend our results to the case of random graph processes with time-varying statistics, we have to perform our analysis in a 2D frequency domain: the graph-frequency domain using the GFT, and the temporal-frequency domain using the Z-transform. Such a type of analysis was previously presented for the deterministic setting in [13], [33], [34].

**FIR.** The following proposition is a generalization of Proposition 1.

*Proposition 2. Consider the FIR$_K$ graph filter (5), and let the graph be RES and the graph process be random with a time-varying statistics. Then, the expected output of the graph filter is*

$$\bar{\boldsymbol{z}}_{t+1} = \sum_{k=0}^{K} \varphi_k \bar{\boldsymbol{L}}^k \bar{\boldsymbol{x}}_{t-k+1}, \tag{9}$$

*which describes a filter on the deterministic time-varying mean signal over the expected graph with the 2-dimensional filter transfer function*

$$H(z,\lambda)=\sum_{k=0}^{K}\varphi_k\left(\frac{\lambda}{z}\right)^k, \quad (10)$$

*where $z$ stems from the Z-transform and $\lambda$ from the GFT.*

From Proposition 2, we can clearly see that the result of Proposition 1 extends to the 2-dimensonal case, capturing jointly the mean signal variations over the expected graph and time. As we will see next, the same result holds also for ARMA graph filters.

**ARMA.** We can now prove the following claim.

*Theorem* 2. *Consider the RES graph model and the random graph process model with time-varying statistics. The relation between the expected filter output and the expected graph signal over the expected graph is given by a 2-dimensional ARMA$_K$ transfer function*

$$H(z,\lambda)=\sum_{k=1}^{K}\frac{\varphi^{(k)}z^{-1}}{1-\psi^{(k)}\lambda z^{-1}}, \quad (11)$$

*subject to the stability conditions of Theorem 1.*

As for the FIR$_K$, Theorem 2 claims that, in expectation, we achieve a 2-dimensional filter operating jointly on the mean variations over the expected graph and time signal.

We conclude this section with the following remarks:

*Remark 1:* Throughout our derivations, we have assumed that the graph realizations $\mathcal{G}_t$ (and thus $\boldsymbol{L}_t$) are independent at different times $t$. This assumption is crucial for our derivations: since $\boldsymbol{z}_t$ is a function of all $\boldsymbol{L}_0,\ldots,\boldsymbol{L}_{t-1}$, when the graphs at different times are dependent $\mathbb{E}[(\boldsymbol{\Psi}\otimes\boldsymbol{L}_t)\boldsymbol{z}_t]\neq(\boldsymbol{\Psi}\otimes\bar{\boldsymbol{L}})\bar{\boldsymbol{z}}_t$, due to $\mathbb{E}[\boldsymbol{L}_{t1}\boldsymbol{L}_{t2}]\neq\mathbb{E}[\boldsymbol{L}_{t1}]\,\mathbb{E}[\boldsymbol{L}_{t2}]$. It is therefore a fact that our results do not generalize to many stochastic graph processes (e.g., human mobility, failure cascade). Our results however do suggest that graph filters are robust to some stochastic phenomena common in networks such as noise and edge fluctuations. Nonetheless, the graph signal $\boldsymbol{x}_t$ might still be a function of the graph topology, e.g., $i)$ the underlying graph $\mathcal{G}$ could be related to $\boldsymbol{\Sigma}_{\boldsymbol{x}}$ or $\boldsymbol{\Sigma}_{\boldsymbol{x}}^{-1}$ by construction, including also the case where $\boldsymbol{\Sigma}_{\boldsymbol{x}}$ shares the graph Laplacian eigenvectors; and $ii)$ as shown in Section III-B the graph signal mean and covariance can depend in an arbitrary way (though deterministic) on the graph topology changes. As a side note, the considered approach with mutual independence between the graph signal and graph topology is also a way to jointly handle the particular cases where only one of them is stochastic.

*Remark 2:* Our second remark is based on the parallelism between stochastic graph filtering and linear system theory [35]. Other works have exploited this analogy to propose graph signal control [36], [37], or graph signal estimation [38], [39]. Differently from linear system theory, the GSP perspective not only gives more insights into what is evolving over the network, but also introduces new strategies related to the characteristics (e.g., smoothness prior, stationarity and bandlimitedness) of the graph signal w.r.t. the underlying graph. More specifically, we can view the ARMA graph filter as a linear state-space model with $\boldsymbol{x}_t$, $\boldsymbol{y}_t=\left[\boldsymbol{y}_t^{(1)\top},\boldsymbol{y}_t^{(2)\top},\cdots,\boldsymbol{y}_t^{(K)\top}\right]^\top$ and $\boldsymbol{z}_t$ being the stochastic input signal, the the $NK\times 1$ stacked state vector of the system and the system output at time $t$, respectively. The stochastic matrix $\boldsymbol{L}_t$, which now captures the stochastic graph topology at time instant $t$, is the stochastic transfer operator of the system. Thus, we can say that the expected behavior of the graph filter is related to the expected behavior of the system and the same holds for the output signal. While in deterministic [40], [41] and stochastic control [42]–[44] of linear systems the focus is on designing the optimal controller, this work considers the analysis/design of the filter transfer function under stochastic variations in the graph topology and/or graph signal.

## IV. Variance Analysis

As we saw in the previous section, when examined in expectation, graph filters are impervious to stochasticity of the graph topology and/or graph signal. To quantify how far from the mean a given realization can be, in the following, we proceed to characterize the covariance of the filter output $\boldsymbol{\Sigma}_{\boldsymbol{z}}$ as a function of that of the input signal $\boldsymbol{\Sigma}_{\boldsymbol{x}}$. Our analysis considers deriving the limiting average variance of the filter output over all nodes, defined as

$$\lim_{t\to\infty}\overline{\text{var}}[\boldsymbol{z}_t]=\lim_{t\to\infty}\frac{tr\left(\boldsymbol{\Sigma}_{\boldsymbol{z}}[t]\right)}{N}=\lim_{t\to\infty}\frac{tr\left(\mathbb{E}[\boldsymbol{z}_t\boldsymbol{z}_t^{\mathsf{H}}]-\bar{\boldsymbol{z}}\bar{\boldsymbol{z}}^{\mathsf{H}}\right)}{N}. \quad (12)$$

In case of FIR graph filters the limit can be omitted. Expression (12), often used for statistical characterizations in signal processing, presents a simple way to quantify the experienced variance of the filter output at each node.

We characterize the variance of the FIR$_K$ and ARMA$_K$ filter output for the general case, where both the graph topology and graph signal change stochastically over time, abiding to a RES graph model and random graph process model, respectively. To the best of our knowledge, finding the variance in closed form is a challenging and open problem not only for graph signal processing, but also in the linear system theory literature. We show that, for both the FIR$_K$ and ARMA$_K$ graph filters, the average variance of the output signal over all nodes (12) is upper bounded. The provided bounds indicate how much a realization of the filtered signal varies around its expected value. Our main goals are: *(i)* to show that the variance of the signal output does not diverge to infinity, and *(ii)* to investigate which parameters influence the bound the most, and as a consequence, which parameter values can lead to a smaller average variance. We note that, as our bounds concern the average variance among all nodes, it is possible that the variance of the signal at a specific node exceeds the bound.

We also present an approach on how to track recursively the variance of the ARMA$_K$ output, i.e., how to calculate the variance of the filter output at time instant $t$, based only on statistical knowledge at the previous time instant. Note that the presented results can also be extended to linear system theory, if the analogies of Remark 2 in Section III-B are used.

**FIR.** The following proposition formally states an upper bound on the average variance of the output signal of the FIR$_K$ filter.

*Proposition* 3. *Consider the FIR$_K$ graph filter* (5) *with a random graph process living on a RES graph. The average variance among all nodes of the FIR$_K$ filter output is bounded by*

$$\overline{var}[\boldsymbol{z}_{t+1}] \leq (\boldsymbol{\varrho}^\top \boldsymbol{\varphi})^2 \left( \overline{var}[\boldsymbol{x}_t] + \frac{\|\bar{\boldsymbol{x}}\|^2}{N} \right), \tag{13}$$

*with vectors $\boldsymbol{\varrho}$ and $\boldsymbol{\varphi}$ respectively, defined as $\boldsymbol{\varrho} = [1, \varrho, \varrho^2, \ldots, \varrho^K]^\top$ and $\boldsymbol{\varphi} = [\varphi_0, \varphi_1, \ldots, \varphi_K]^\top$.*

The result of Proposition 3 suggests that despite the drastic variation of the graph topology, the average variance of the output signal over the nodes is finite and thus also the variance of the output signal at a given node. We can see that the filter order $K$ has indeed an impact on the bound.

**ARMA.** As for the FIR, the following theorem gives an expression on how to upper bound the average variance across all nodes of the ARMA graph filter output.

*Theorem* 3. *Consider the ARMA$_K$ graph filter* (7) *with a random graph process living on a RES graph and additionally assume that the initial filter state $\boldsymbol{y}_0$ is independent of $\boldsymbol{x}_t$ with zero mean $\mathbb{E}[\boldsymbol{y}_0] = \boldsymbol{0}$. The limiting average variance among all nodes of the ARMA$_K$ filter output is bounded by*

$$\lim_{t\to\infty} \overline{var}[\boldsymbol{z}_{t+1}] \leq \frac{K\|\boldsymbol{\varphi}\|^2}{(1-\varrho\,|\psi_{max}|)^2} \left( \overline{var}[\boldsymbol{x}_t] + \frac{\|\bar{\boldsymbol{x}}\|^2}{N} \right) \tag{14}$$

*where $\psi_{max} = \max\{\psi^{(1)}, ..., \psi^{(K)}\}$.*

According to Theorem 3, the bound depends linearly on the filter order $K$ and on the sum of the squared coefficients $\boldsymbol{\varphi}$.

We now continue with the recursive approach to exactly calculate the variance of the filter output. We directly consider the case of graph process with time-varying statistics, since, differently from the bounds (13)-(14), the recursive variance is less challenging to compute. Recalling that the graph Laplacian at time $t$, i.e., $\boldsymbol{L}_t$, is a random realization with mean $\bar{\boldsymbol{L}}$, it can be written as $\boldsymbol{L}_t = \bar{\boldsymbol{L}} + \tilde{\boldsymbol{L}}_t$, with $\tilde{\boldsymbol{L}}_t$ being the related zero-mean process. We then define the matrices

$$\bar{\boldsymbol{A}} = \boldsymbol{\Psi} \otimes \bar{\boldsymbol{L}},\ \tilde{\boldsymbol{A}}_t = \boldsymbol{\Psi} \otimes \tilde{\boldsymbol{L}}_t \text{ and } \boldsymbol{B} = \boldsymbol{\varphi} \otimes \boldsymbol{I}_N \tag{15}$$

and rewrite the ARMA$_K$ filter (7) as

$$\boldsymbol{y}_{t+1} = \bar{\boldsymbol{A}}\boldsymbol{y}_t + \tilde{\boldsymbol{A}}_t\boldsymbol{y}_t + \boldsymbol{B}\boldsymbol{x}_t \tag{16a}$$

$$\boldsymbol{z}_{t+1} = \boldsymbol{C}\boldsymbol{y}_{t+1}, \tag{16b}$$

with $\boldsymbol{\Psi} = \mathrm{diag}(\psi^{(1)}, \psi^{(2)}, \cdots, \psi^{(K)})$ a diagonal $K \times K$ coefficient matrix and $\boldsymbol{C} = \boldsymbol{1}^\top \otimes \boldsymbol{I}_N$ with $\boldsymbol{1}$ a $K \times 1$ vector. The instantaneous expected value of the output signal of (16) is calculated as

$$\bar{\boldsymbol{y}}_{t+1} = \bar{\boldsymbol{A}}\bar{\boldsymbol{y}}_t + \boldsymbol{B}\bar{\boldsymbol{x}}_t \tag{17a}$$

$$\bar{\boldsymbol{z}}_{t+1} = \boldsymbol{C}\bar{\boldsymbol{y}}_{t+1}, \tag{17b}$$

which suggests a way to recursively track the expected value of the output signal at time instant $t+1$ based on the knowledge of the first order statistics at the previous time instant $t$. With this in place, we can state the following

*Proposition* 4. *Given the ARMA$_K$ graph filter* (16) *operating on a random graph process with time-varying statistics over a RES graph model and given first and second order moments of the filter output at time instant $t$. Then, the covariance matrix of the output signal at time instant $t+1$ can be calculated as*

$$\begin{aligned} \boldsymbol{\Sigma_z}[t+1] = &\ \boldsymbol{C}(\bar{\boldsymbol{A}}\boldsymbol{\Sigma_y}[t]\bar{\boldsymbol{A}}^\top + \boldsymbol{B}\boldsymbol{\Sigma_x}[t]\boldsymbol{B}^\top)\boldsymbol{C}^\top \\ &+ \boldsymbol{C}\mathbb{E}_{\tilde{\boldsymbol{A}}}\left[\tilde{\boldsymbol{A}}_t\boldsymbol{R_y}[t]\tilde{\boldsymbol{A}}_t^\top\right]\boldsymbol{C}^\top \end{aligned} \tag{18}$$

*where $\boldsymbol{R_y}[t] = \mathbb{E}_{\boldsymbol{y}}\left[\boldsymbol{y}_t\boldsymbol{y}_t^\top\right]$ with $\mathbb{E}_{\tilde{\boldsymbol{A}}}[\cdot]$ the expected value only with respect to the random variable $\tilde{\boldsymbol{A}}$.*
*The entry $(i,j)$ of $\mathbb{E}_{\tilde{\boldsymbol{A}}}[\tilde{\boldsymbol{A}}_t\boldsymbol{R_y}[t]\tilde{\boldsymbol{A}}_t^\top]$ can be calculated as*

$$\begin{aligned} \mathbb{E}_{\tilde{\boldsymbol{A}}}\left[\tilde{\boldsymbol{A}}_t\boldsymbol{R_y}[t]\tilde{\boldsymbol{A}}_t^\top\right]_{i,j} &= \sum_{n=1}^{N} C_{A_{j1},A_{in}}\left[\boldsymbol{R_y}[t]\right]_{n,1} \\ &+ \sum_{n=1}^{N} C_{A_{j2},A_{in}}\left[\boldsymbol{R_y}[t]\right]_{n,2} + \ldots + \sum_{n=1}^{N} C_{A_{jN},A_{in}}\left[\boldsymbol{R_y}[t]\right]_{n,N}, \end{aligned} \tag{19}$$

*with $C_{A_{oq},A_{rs}} = cov(A_{oq}[t], A_{rs}[t])$ the covariance between the $(o,q)$-th and $(r,s)$-th entries of the matrix $\tilde{\boldsymbol{A}}_t$, which can be expressed as a function of the link activation probability and filter coefficients.*

The result of Propositon 4 gives a way to track recursively the variance of the filter output signal, and thus to stochastically characterize how far a given realization may be at each node. Further, as expected, the lower the variance of the input signal the lower the variance of the output.

We conclude this section with the following remarks.

*Remark 3:* We see from both bounds and the recursive variance, that the order of the filter and the sum of the squared coefficients have an influence on the average variance. Thus, our handle on obtaining a small variance are the filter order and the coefficient values. The latter can be a problem for the FIR filter, where the magnitude of the coefficients is much greater than for the ARMA filter. Further, their value tends to increase with the filter order. One way to reduce the filter coefficients value and therefore the variance is to work with the shifted Laplacian $\boldsymbol{L} = \boldsymbol{L}_{\mathrm{d}} - \lambda_{max}/2\boldsymbol{I}$, or $\boldsymbol{L} = \boldsymbol{L}_{\mathrm{n}} - \boldsymbol{I}$, instead of using directly the discrete Laplacian $\boldsymbol{L}_{\mathrm{d}}$, or normalized Laplacian $\boldsymbol{L}_{\mathrm{n}}$. Also, considering that the spectral norm of $\boldsymbol{L}$ (i.e., $\varrho$) has an impact on the bounds (13) - (14), the use of Laplacian matrices with small spectral norms, such as the ones mentioned above, is recommended to achieve a small variance in the output signal.

*Remark 4:* For the particular case where only the graph signal is random time-varying, living on a deterministic time-invariant graph, the variance can be calculated in closed form in complete analogy with linear system theory and using the derivations for the ARMA$_1$ in [18].

*Remark 5:* While the above derivations concern the graph filter behavior in a stochastic environment, one could extend the analysis in the properties of random graph signals over random graphs. Thus, for a deterministic graph signal $\boldsymbol{x}$ living on a random graph $\mathcal{G}_t$, the 2-Dirichlet form (Df$_2$) defined as $\mathrm{Df}_2 = \boldsymbol{x}^\mathsf{T}\boldsymbol{L}\boldsymbol{x}$ [5] will be a random variable, with mean $\mathbb{E}[\mathrm{Df}_2] = \boldsymbol{x}^\mathsf{T}\bar{\boldsymbol{L}}\boldsymbol{x}$ and variance $var(\mathrm{Df}_2) = \mathbb{E}[(\boldsymbol{x}^\mathsf{T}(\boldsymbol{L}-\bar{\boldsymbol{L}})\boldsymbol{x})^2]$. In this setting, smooth graph signals over random graphs are characterized by being smooth in the mean and by having $var(\mathrm{Df}_2)$ as small as possible. Similarly when $\boldsymbol{x}$ is a random process, the reasoning extends by incorporating also the statistics of $\boldsymbol{x}$ in the $\mathrm{Df}_2$ analysis.

## V. Leveraging Stochasticity

So far, we have analyzed temporal stochasticity on the graph topology and graph signal as problems that can affect the filter behavior. In this section, we view the stochasticity from another perspective. We present two approaches that exploit stochasticity: one to improve the quality of denoising when multiple realizations of the graph signal are available, and one to estimate the output of a graph filter with a graph filter of lower communication complexity.

### A. Graph signal denoising in the mean

Consider a noisy graph signal of the form $\boldsymbol{x} = \boldsymbol{u} + \boldsymbol{n}$, with $\boldsymbol{u}$ the signal of interest and $\boldsymbol{n}$ the zero mean additive noise. The goal is to recover $\boldsymbol{u}$ using the a-priori knowledge on its behavior w.r.t. the underlying graph.

**Tikhonov denoising**. Tikhonov denoising recovers the signal of interest $\boldsymbol{u}$ by exploiting the smoothness prior of the latter w.r.t. the underlying graph [5] (and references therein). For instance, this can be the case of temperature measurements, where adjacent nodes share similar values. The solution for $\boldsymbol{u}$ can be cast as

$$\boldsymbol{u}^* = \underset{\boldsymbol{u}\in\mathbb{R}^N}{\mathrm{argmin}} \|\boldsymbol{u} - \boldsymbol{x}\|_2^2 + w\, \boldsymbol{u}^\mathsf{T}\boldsymbol{L}\boldsymbol{u}, \tag{20}$$

where the regularizer $\boldsymbol{u}^\mathsf{T}\boldsymbol{L}\boldsymbol{u}$ is the prior assumption that the graph signal varies smoothly on the graph which is weighted by the constant $w$. In (20), admitted choices of $\boldsymbol{L}$ are limited to the discrete Laplacian $\boldsymbol{L}_\mathrm{d}$ or the normalized Laplacian $\boldsymbol{L}_\mathrm{n}$ (without shifting). Under this prior, problem (20) attains the closed form solution

$$\boldsymbol{u}^* = (\boldsymbol{I} + w\boldsymbol{L})^{-1}\boldsymbol{x} = \sum_{n=1}^{N} \frac{1}{1 + w\lambda_n} \langle \boldsymbol{x}, \boldsymbol{\phi}_n \rangle \boldsymbol{\phi}_n. \tag{21}$$

As can be seen from (21), the optimal solution $\boldsymbol{u}^*$ can be obtained through filtering of $\boldsymbol{x}$ with a rational frequency response filter. Then, from the results of [13], an $\mathrm{ARMA}_1$ filter can be used to exactly find $\boldsymbol{u}^*$. In this case, the task can be solved without knowing the full graph topology since: $i$) the filter coefficients can be found avoiding the eigendecomposition, i.e., only the value of $w$ is necessary; and $ii$) in a distributed implementation nodes just need to know who are their neighbors to run the filter.

**Denoising in the mean.** When multiple realizations of $\boldsymbol{x}$ are available (e.g., in a sensor network) one may consider to generalize (20) so as to improve the accuracy of the denoising problem. In this case, we have for every realization $\boldsymbol{x}_t = \boldsymbol{u} + \boldsymbol{n}_t$ with $\boldsymbol{u}$ equal to the desired signal. The graph signal denoising problem can then be performed *in the mean.* Similar to (20), the related optimization problem is

**Algorithm 1** *Disjoint average and denoise*

1: Given $w$, initialize $\psi$, $\varphi$, $\boldsymbol{y}_0 = \boldsymbol{0}$ and the $\mathrm{ARMA}_1$ iterations $I$
2: **for** each $t > 0$
3: *compute local average* $\bar{x}_{t,i} = \frac{1}{t}((t-1)\bar{x}_{t-1,i} + x_{t,i})$
4: **procedure** Compute the estimate of $u^*_{t,i}$
5:     **for** $\iota = 1$ **to** $I$
6:     *collect* $y_j[\iota-1]$ *from all neighbors* $j \in \mathcal{N}_i$
7:     *compute* $y_i[\iota] = \psi \sum_{j\in\mathcal{N}_i} L_{i,j}\,(y_i[\iota-1] - y_j[\iota-1]) + \varphi\bar{x}_{t,i}$
8:     *send* $y_i[\iota]$ *to all neighbors* $\mathcal{N}_i$
9:     **end for**
10:     *set* $u^*_{t,i} = y_i[I]$
11: **end procedure**
12: **end for**

$$\boldsymbol{u}^* = \underset{\boldsymbol{u}\in\mathbb{R}^N}{\mathrm{argmin}}\, \mathbb{E}\Big[\|\boldsymbol{u} - \boldsymbol{x}_t\|_2^2\Big] + w\, \boldsymbol{u}^\mathsf{T}\boldsymbol{L}\boldsymbol{u}, \tag{22}$$

where we aim to find the best solution in the mean squared error sense over different time realizations. Analogous to the solution of (21), the optimal solution of (22) can still be written as an $\mathrm{ARMA}_1$ filter of the form

$$\boldsymbol{u}^* = \sum_{n=1}^{N} \frac{1}{1 + w\lambda_n} \langle \mathbb{E}[\boldsymbol{x}_t]\,, \boldsymbol{\phi}_n \rangle \boldsymbol{\phi}_n, \tag{23}$$

which has two interpretations: (i) it can be considered the output of an $\mathrm{ARMA}_1$ graph filter applied to the mean $\bar{\boldsymbol{x}}_t$ or (ii) the mean output of an $\mathrm{ARMA}_1$ applied to $\boldsymbol{x}_t$.
In this regard, the time-graph denoising can be performed as:

*i) Disjoint average and denoise (DAD).* One approach for solving (23) is to first perform a running average up to time $t$, independently at each node $\bar{x}_{t,i} = \frac{1}{t}((t-1)\bar{x}_{t-1,i} + x_{t,i})$, and then to run a recursive $\mathrm{ARMA}_1$ on the average signal $\bar{\boldsymbol{x}}_t = [\bar{x}_{t,1}, \ldots, \bar{x}_{t,N}]^\mathsf{T}$. Hence, these recursions are not in time, but they are carried out for every $t$ separately. This approach is summarized in Algorithm 1. Since the noise is zero mean the local average [cf. line 3 of Algorithm 1] reduces the noise level proportionally to the number of collected samples, effectively facilitating the work of graph denoising. However, computationally this algorithm needs to run the graph filter till convergence for each available sample of the input signal $\boldsymbol{x}_t$. We refer to this approach as *disjoint average and denoise*.

*ii) Joint denoise in the mean.* The complexity of denoising in the mean can be reduced by performing the local averaging and the denoising jointly, an approach which we refer to as *joint denoising in the mean*. In virtue of the results in Section III, we distinguish between three different cases to perform the local averaging.

$ii - a$) The more general case of joint denoising in the mean with *input-output averaging* (JDMIOA), where a regular ARMA filter[2] is run in time with as input signal at time $t$ the local *running average* of the past history of $\boldsymbol{x}_t$ and as the final

[2]From [13], the $\mathrm{ARMA}_1$ enjoys a stable implementation for all $w > 0$.

**Algorithm 2** *Joint denoise in the mean with input-output averaging (JDMIOA)*

```
1: Given w, initialize ψ, φ and y_0 = 0
2: for each t > 0
3: compute local average x̄_{t,i} = 1/t((t − 1)x̄_{t−1,i} + x_{t,i})
4: procedure COMPUTE THE ESTIMATE OF u*_{t,i}
5:     collect y_{t−1,j} from all neighbors j ∈ N_i
6:     compute y_{t,i} = ψ Σ_{∈N_i} L_{i,j} (y_{t−1,i} − y_{t−1,j}) + φx̄_{t,i}
7:     send y_{t,i} to all neighbors N_i
8:     set u*_{t,i} = 1/t((t − 1)u*_{t−1,i} + y_{t,i})
9: end procedure
10: end for
```

solution of (23) at node $i$ the local *running average* of the filter output up to time $t$, i.e., $u^*_{t,i} = 1/t((t-1)u^*_{t-1,i} + y_{t,i})$. This procedure is summarized in Algorithm 2.

$ii{-}b$) The particular case of joint denoising in the mean with *input averaging* (JDMIA), which differently from JDMIOA sets the final solution of (23) at node $i$ as the filter output at time $t$, i.e., in Algorithm 2 this differs only in line 8 which should be "*set* $u^*_{t,i} = y_{t,i}$". Note that this procedure is a direct online implementation of the DAD algorithm.

$ii - c$) The particular case of joint denoising in the mean with *output averaging* (JDMOA), which differently from the JDMIOA has as input signal at time $t$ the collected samples, $\boldsymbol{x}_t = \boldsymbol{u} + \boldsymbol{n}_t$. Referring again to Algorithm 2, this differs only in line 3 which should be "*collect the local sample* $x_{t,i} = u_i + n_{t,i}$".

Despite being particular cases of JDMIOA, the JDMIA and JDMOA perform differently depending on the scenario. Note that the joint denoising in the mean approaches follow the same filtering strategy as a classical graph signal denoising filter, and thus do not require extra memory, computations or communication efforts. Again, the noise being zero mean, averaging the filter input and/or output reduces the noise power that will be introduced by the filter in every successive iteration, while performing the online denoising in the mean.

$iii$) *Local averaging (LA).* As a benchmark we will consider also the local averaging, i.e., the nodes ignore the presence of the graph and they directly set $\boldsymbol{u}^*_t = 1/t((t-1)\boldsymbol{u}^*_{t-1} + \boldsymbol{x}_t)$.

Numerical results show that the joint denoise in the mean algorithms outperform pure local averaging when the noise level is high and when the number of measurements are limited. Further, as the number of available samples increases, the joint algorithms ensure the same performance as the DAD algorithm at a smaller computational cost. We conclude this section stressing the fact that our ideas are not restricted to denoising or simple ARMA$_1$ filtering. It can be used for any filter, ARMA$_K$ or FIR$_K$, as long as we are able to make multiple observations of a signal corrupted with additive zero mean noise.

### *B. Stochastically sparsified graph filtering*

To decrease the communication and computational complexity of graph filtering, we propose, instead of filtering a deterministic signal $\boldsymbol{x}$ on the deterministic graph $\mathcal{G}$, to filter $\boldsymbol{x}$ using realizations of a sparser random graph $\mathcal{G}_t$ obtained as a random edge sampling of $\mathcal{G}$. This means that the filtering will be performed on a sequence of time-varying graphs which abide to the RES graph model with a uniform probability $p \in (0, 1]$ for all links. What we prove in the following is that, when the filter coefficients are modified accordingly, the filter output in $\mathcal{G}_t$ will be statistically close (the first and second moments of the error are zero and bounded, respectively) to the original filter in $\mathcal{G}$. Our results imply a reduction in the communication and computational complexity of graph filtering that is linear in $p$.

Denote by $\boldsymbol{L}$ the Laplacian of the deterministic graph $\mathcal{G}$, where (for now) $\boldsymbol{L}$ is chosen as the discrete Laplacian $\boldsymbol{L}_\text{d}$.

**FIR.** With a slight modification of our results in Section III (since the graph signal is now deterministic), we find that in expectation the filter output will be

$$\bar{\boldsymbol{z}}^{(\text{s})}_{t+1} = \mathbb{E}\Big[\boldsymbol{z}^{(\text{s})}_{t+1}\Big] = \mathbb{E}\left[\sum_{k=0}^{K} \varphi^{(\text{s})}_k \boldsymbol{\Phi}_{\boldsymbol{L}}(t, t-k+1)\boldsymbol{x}\right] = \sum_{k=0}^{K} \varphi^{(\text{s})}_k \bar{\boldsymbol{L}}^k \boldsymbol{x} = \sum_{k=0}^{K} \varphi^{(\text{s})}_k (p\boldsymbol{L})^k \boldsymbol{x}, \quad (24)$$

where $\boldsymbol{z}^{(\text{s})}_{t+1}$ is the stochastic sparsification output and $\varphi^{(\text{s})}_k$ are the new coefficients. Therefore, if each coefficient $\varphi^{(\text{s})}_k$ is $\varphi_k$ scaled by $p^{-k}$, the expected output will be identical to what would have been obtained if the original FIR filter was used in $\mathcal{G}$

$$\mathbb{E}\Big[\boldsymbol{z}^{(\text{s})}_{t+1}\Big] = \boldsymbol{z}_{t+1} \quad \text{for} \quad \varphi^{(\text{s})}_k = \varphi_k p^{-k}, \quad (25)$$

or equivalently, the expected sparsification error is zero

$$\mathbb{E}\Big[\boldsymbol{z}^{(\text{s})}_{t+1} - \boldsymbol{z}_{t+1}\Big] = \boldsymbol{0}. \quad (26)$$

The above expression implies that we can linearly reduce the communication and computational complexity of FIR graph filters from $O(MK)$ to $O(pMK)$. To see how $p$ influences how far the error can lie from its mean, in the following, we derive an upper bound on the average variance of the output signal. Directly from Proposition 3, we have that

$$\overline{\text{var}}[\boldsymbol{z}^{(\text{s})}_{t+1}] \leq (\boldsymbol{\varrho}^\top \boldsymbol{\varphi}^{(\text{s})})^2 \left(\overline{\text{var}}[\boldsymbol{x}] + \frac{\|\bar{\boldsymbol{x}}\|^2}{N}\right) = (\boldsymbol{r}^\top \boldsymbol{\varphi})^2 \frac{\|\boldsymbol{x}\|^2}{N}, \quad (27)$$

with $\boldsymbol{r} = [(\varrho/p)^0, (\varrho/p)^1, \ldots, (\varrho/p)^K]^\top$. Though the provided bound is not tight (as is witnessed by the fact that for $p = 1$ the variance is not zero), it illustrates the impact of $1/p$ on the error variance.

**ARMA.** As for the FIR, we can derive that the expected sparsified output $\bar{\boldsymbol{z}}^{(\text{s})}_{t+1}$ of the ARMA$_K$ filter is

$$\begin{aligned} \bar{\boldsymbol{y}}^{(\text{s},k)}_{t+1} &= \psi^{(\text{s},k)}(p\boldsymbol{L})\bar{\boldsymbol{y}}^{(\text{s},k)}_t + \varphi^{(k)}\boldsymbol{x} \\ \bar{\boldsymbol{z}}^{(\text{s})}_{t+1} &= \sum_{k=1}^{K} \bar{\boldsymbol{y}}^{(\text{s},k)}_{t+1}, \end{aligned} \quad (28)$$

where again the superscript (s) refers to the sparsified filtering. We now notice, that as long as $\psi^{(\text{s},k)} = \psi^{(k)}/p$ the expected output of (28) is identical to the same ARMA filter operating on the complete graph. This sparsification will again reduce the communication and computational complexity by the same order as for the FIR. Further, from Theorem 3, the average variance of the sparsified filter can now be upper bounded by

$$\lim_{t\to\infty} \overline{\text{var}}[\boldsymbol{z}_{t+1}^{(\text{s})}] \leq \frac{K\|\boldsymbol{\varphi}\|^2\|\boldsymbol{x}\|^2}{N(1-\varrho\,|\frac{\psi_{max}}{p}|)^2} \tag{29}$$

As for the FIR, the bound (29) does not reach zero for $p = 1$, but it indicates that the variance is upper bounded by a quadratic rate of $1/p$.

We conclude this section with the following observations.

*Remark 6:* Our numerical simulations show that, even though in practice the sparsification error variance is much smaller than what is claimed by (27) and (29), $p$ has indeed the claimed impact on the output variance.

*Remark 7:* The above theoretical derivations provide a useful way to modify the filter coefficients, in order to avoid a bias in the sparsified approach. These derivations stand only for the discrete Laplacian $\boldsymbol{L} = \boldsymbol{L}_\text{d}$ (or its translated version $\boldsymbol{L} = \boldsymbol{L}_\text{d} - \boldsymbol{I}$), meaning that the expected graph Laplacian can be expressed as a scaled version of the Laplacian of the underlying graph $\mathcal{G}$. This is not the case for the normalized Laplacian $\boldsymbol{L}_\text{n}$ (or its translated version $\boldsymbol{L} = \boldsymbol{L}_\text{n} - \boldsymbol{I}$). Finding a closed form expression of the mean normalized Laplacian is a challenging task. Even if it can be found, its $(i, j)$-th entry will depend on the node degrees in $\mathcal{G}$, which renders formulating $\bar{\boldsymbol{L}}_\text{n}$ as a scaled version of the normalized Laplacian $\boldsymbol{L}_\text{n}$ difficult. However, the use of the (translated) normalized Laplacian is useful to improve the stability region of the ARMA filter and also in some cases it provides FIR coefficients with lower magnitude (and as a consequence lead to a lower error variance). As we will see in the simulation results, the use of $\boldsymbol{L} = \boldsymbol{L}_\text{n} - \boldsymbol{I}$, even *without changing the filter coefficients*, will lead to an error with practically a very small mean and variance.

## VI. Numerical Results

For our simulation results, we consider a graph of $N = 100$ nodes randomly placed in a square area and two nodes are considered neighbors if they are closer than 15% of the maximum distance of this area. The comparison of the filter output is done at steady-state, i.e., for the FIR the output is considered after $K$ time instants, while for the ARMA after $20 \times K$ iterations and its initial state is $\boldsymbol{y}_0 = \boldsymbol{0}$. The empirical results are averaged over 2000 simulation runs. If not explicitly mentioned, in our simulations, we will run the filters w.r.t. the translated normalized Laplacian $\boldsymbol{L} = \boldsymbol{L}_\text{n} - \boldsymbol{I}$. Such a choice has the benefit to improve the ARMA stability region [13] and produces FIR coefficients of lower magnitude.

### A. Filtering graph processes

In our first scenario, we quantify the behavior of the graph filters under stochasticity. Our goal is to show that graph filters can handle stochastic variations of the graph topology and graph signal. This scenario can be considered as the case when we want to perform the filtering on a deterministic graph with a deterministic signal, but we have only random realizations of the latter. We also aim to quantify the variance of the output signal, both empirically and with upper-bounds. We analyze both FIR and ARMA filters for $K = 1, 5, 10$ applied to an input signal of the form $\boldsymbol{x}_t = \bar{\boldsymbol{x}} + \boldsymbol{n}_t$, with

$$\langle \bar{\boldsymbol{x}}, \boldsymbol{\phi}_n \rangle = \begin{cases} 1 & \text{if } \lambda_n < 1 \\ 0 & \text{if } \lambda_n \in [1, 2] \end{cases} \tag{30}$$

being low pass, and noise $\boldsymbol{n}_t$ which follows a zero mean normal distribution with covariance matrix $\boldsymbol{\Sigma}_n = \sigma_n^2 \boldsymbol{I}$. All graph filters (FIR and ARMA) are designed to approximate a frequency response identical to the one of the graph signal $\bar{\boldsymbol{x}}$. We analyze the filtering performance for different link-activation probabilities $p$ and noise powers $\sigma_n^2$. We compare the filtered signal under the stochastic model $\boldsymbol{z}^{(\text{s})}$ with the deterministic output signal $\boldsymbol{z}^{(\text{d})}$ of the same filters operating on the expected graph with graph signal $\bar{\boldsymbol{x}}$. The expected (translated) normalized Laplacian is estimated over 1000 runs.

To quantify the performance we define the error

$$\boldsymbol{e} = \boldsymbol{z}^{(\text{s})} - \boldsymbol{z}^{(\text{d})}. \tag{31}$$

Given the zero mean-error (31), our conclusions will be based on the empirical average standard deviation among all nodes and realizations

$$\bar{\sigma}_{\boldsymbol{e}} = \left[tr\left(\mathbb{E}\Big[\boldsymbol{e}\boldsymbol{e}^{\mathsf{H}}\Big]\right)/N\right]^{\frac{1}{2}} = \left[tr\Big(\boldsymbol{\Sigma}_{\boldsymbol{z}_\text{s}}\Big)/N\right]^{\frac{1}{2}}, \tag{32}$$

which for a sufficiently large number of realizations approaches the square root of the average variance used in our bounds, i.e., $\bar{\sigma}_{\boldsymbol{e}} \approx \sqrt{\overline{\text{var}}[\boldsymbol{z}_s]}$. Fig. 1 depicts the empirical average standard deviation for different values of $p$ and $\sigma_n^2$. We make the following observations: *First, ARMA graph filters yield smaller variance for low filter orders ($K = 1$ and 5), as compared to FIR.* Though the results are not included here due to space constraints, the order under which ARMA filters are preferable to FIR is $K = 8$. The increased variance of high order ARMA filters examined here is mostly related to their higher convergence time, thus an higher error is introduced in the filter memory through $\boldsymbol{y}_t$ while computing the "steady state" output. The effect could be ameliorated by imposing a stricter stability constraint in the design problem, asking now that $|p_k| \geq c > 1$ for some constant $c$ in (4), which is translated in a faster convergence at the expense of a worse filter approximation. *Second, there is an exponential dependence between the output and input variance.* This phenomenon becomes clear when the noise is the only source of variance ($p = 1$). The results in Fig. 1 suggest that graph filters can attain a reasonable performance with stochastically time-varying signals as long as the noise variance is lower than $\sigma_n^2 < 10^{-1}$. However, as shown in Section VI-B, filtering a stochastic signal can have a beneficial outcome, for example in the case of denoising, if compared not to the deterministic filter output $\boldsymbol{z}^{(\text{d})}$ but to the denoising solution. For $\sigma_n^2 \geq 10^{-3}$, we notice that $p$ has a smaller influence on $\bar{\sigma}_{\boldsymbol{e}}$. *Last, we observe that lower-order filters (especially ARMA) tolerate a significant amount of edge fluctuation when the noise variance*

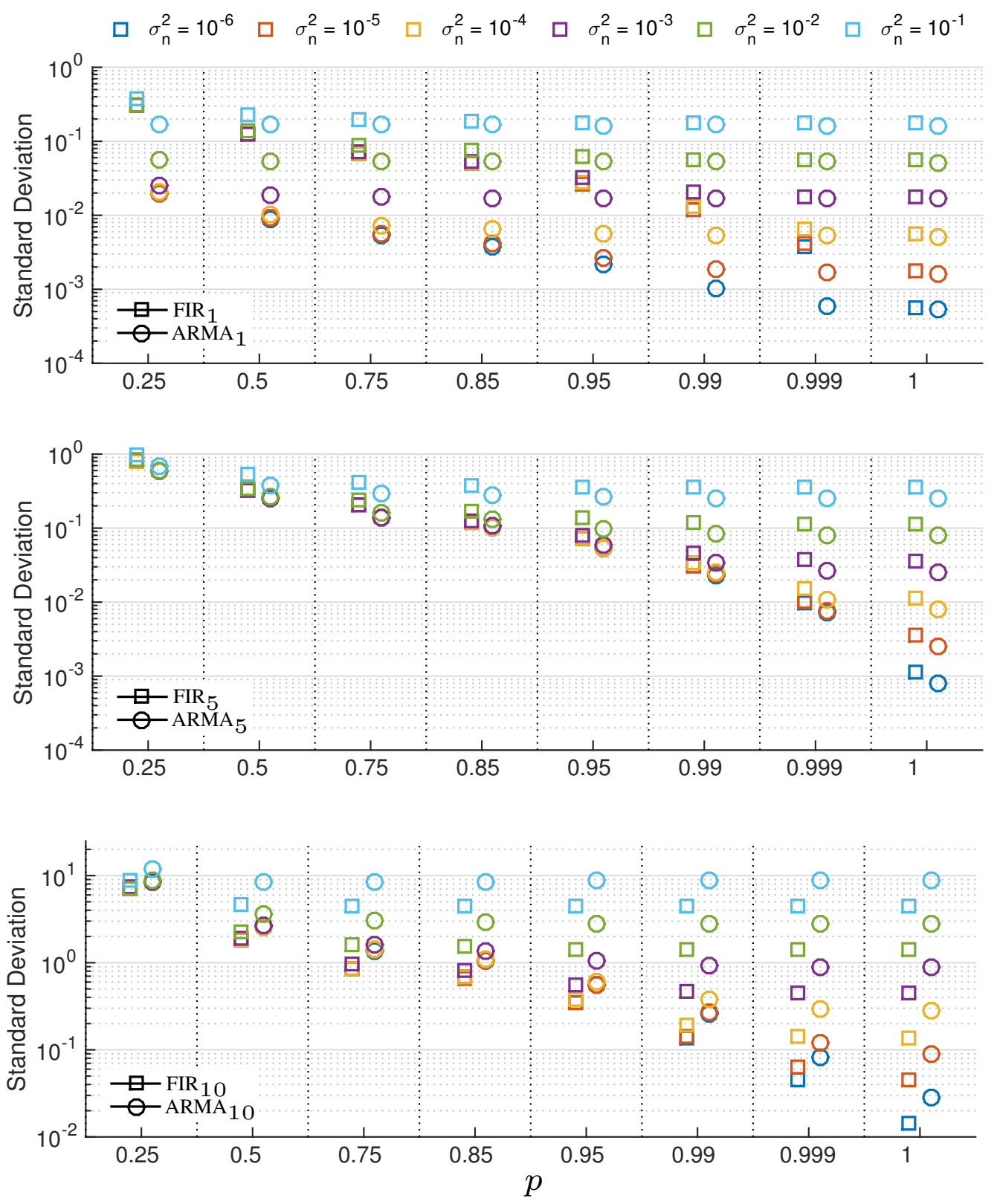

Fig. 1: Empirical average standard deviation among all nodes and all realizations vs. the link activation probability $p$ and for different noise levels $\sigma_n^2$. From top to bottom we have K = 1, 5 and 10, respectively. ARMA filters seem to handle stochasticity better for low orders, while FIR filters perform better for K = 10.

*is small.* This finding motivates the use of graph filters in stochastic time-varying scenarios, as occurring in distributed systems with message loss. A more comprehensive analysis is presented in the context of stochastic sparsification in Section VI-C.

Table I depicts the square roots of the upper bounds (13) and (14) for the FIR and ARMA filters, respectively. The bounds are shown here for low filter orders, since for higher values of $K$ they become increasingly less tight. Despite the difference between the variance guaranteed by the upper bounds and the empirical variance, the results in Table I and Fig. 1 suggest that *(i)* the filter order has an impact on the actual variance as well and *(ii)* in practice the actual variance is much lower.

### B. Denoising in the mean

We now illustrate the denoising in the mean problem of Section V-A and compare the performance of all the joint approaches with both the DAD and the LA approach. For our scenario, we consider a graph signal $\boldsymbol{x}_t = \boldsymbol{u} + \boldsymbol{n}_t$, where the signal of interest $\boldsymbol{u}$ varies smoothly over the graph. One way to generate these signals is to consider an exponentially decaying spectrum, i.e., $\langle \boldsymbol{u}, \boldsymbol{\phi}_n \rangle = \exp(-25\lambda_n)$. The noise follows a zero mean normal distribution with i.i.d. realizations and variance $\sigma_n^2 = 1$. To quantify the performance, we calculate the error between the filter output and the signal $\boldsymbol{u}$ at each node as in (32).

TABLE I: Square roots of (13) / (14) for the FIR / ARMA filters.

| *order* | $\sigma_n^2 = 10^{-6}$ | $\sigma_n^2 = 10^{-3}$ | $\sigma_n^2 = 10^{-1}$ |
|---|---|---|---|
| K=1 | 0.88 / 0.58 | 0.88 / 0.58 | 0.93 / 0.6 |
| K=3 | 0.58 / 6.9 | 0.58 / 6.9 | 0.66 / 7.2 |
| K=5 | 0.82 / 22.8 | 0.82 / 22.8 | 0.88 / 23.9 |

There is a subtle yet crucial difference between the inner-workings of the three compared algorithms: LA and joint denoising in the mean are online algorithms, meaning that the algorithmic output is obtained in real-time[3]. The error depicted in the figures at time $t$ is therefore computed w.r.t. the denoising output at $t$. On the other hand, the DAD algorithm performs the averaging and denoising steps in sequence, and subsequently solves one filtering problem for each computed local average (i.e., at each iteration $t$). Since each disjoint filtering requires sufficient iterations to converge, the error depicted in the figure at $t$ is in this case computed w.r.t. the filter output at convergence, occurring much later than $t$ (we took $t + 100$ in our experiments, but also $t + 10$ can be considered as a choice).

**Tikhonov denoising in the mean.** In Fig. 2(a) we compare the LA (solid black line), the DAD and all three joint denoising in the mean approaches as a function of time for Tikhonov denoising. In this case $w$ is 0.5. The first thing to notice is that incorporating the graph knowledge via the smoothness outperforms the LA. Specifically, the role of the graph is more important when less samples are available, i.e., small values of $t$. Secondly, while the JDMIA immediately approaches the DAD (being it an online implementation of the latter), the JDMIOA and JDMOA for $t \leq 10$ perform better. Then, when more samples become available, the JDMOA matches the DAD. On the other hand, the JDMIOA prioritizes the noise reduction over the smoothness prior due to the double averaging of the filter input and output (lines 3 and 8 in Algorithm 2, respectively). Thus, when more samples are available the related $\bar{\sigma}_e$ decays as that of the LA algorithm. To avoid overcrowded plots, in the sequel, we don't show the performance of the JDMIA, since we consistently observed that it directly matches the DAD algorithm.

Then, in Fig. 2(b) we show the performance of the JD-MIOA, JDMOA and DAD algorithm when the weight given to the smoothness prior is twice that of the denoising, i.e., $w = 2$. As we can see, taking into account the graph structure (i.e., $w$ is higher w.r.t. Fig. 2(a)) improves the performance as it exploits the smoothness prior to remove the noise. Once again, when $t$ increases the JDMOA approaches the DAD algorithm, while the JDMIOA offers a performance between the latter

[3]In our experiments, the signal changes with the same rate as our iteration rate. For this reason, we use the term iteration or more simply time, to refer to both concepts.

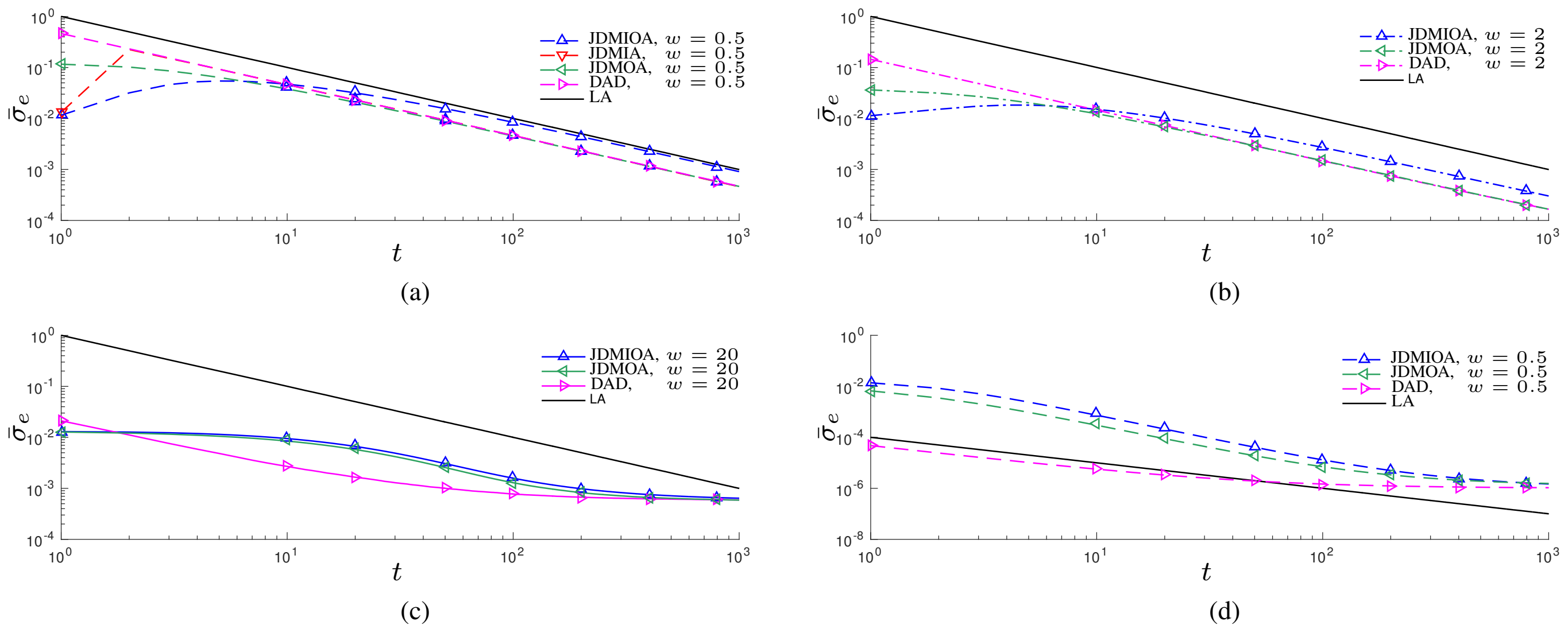

Fig. 2: Standard deviation of the different denoising algorithms for different time instances. Fig. 2(a) compares the joint denoising in the mean algorithms with the DAD and LA algorithms for $w = 0.5$ and $\sigma^2 = 1$. Fig.2(b) shows the performance for $w = 2$ (i.e., where smoothness is prioritised over denoising) and $\sigma_n^2 = 1$. Fig.2(c) compares the algorithms for $w = 20$ and $\sigma_n^2 = 1$. Finally, in Fig.2(d) is shown the error standard deviation for $w = 0.5$ and $\sigma_n^2 = 10^{-4}$.

and the LA.

Fig. 2(c) compares the algorithms for $w = 20$. In this case, when a large number of samples is available, a large $w$ may overly prioritize the graph smoothness resulting in a performance degradation. Since in this instance the noise becomes negligible, the smoothness prior is only useful when the number of samples is limited ($t < 100$). This is not only an issue of our joint approach, but of graph denoising in general.

This effect is also illustrated in Fig. 2(d) for $\sigma_n^2 = 10^{-4}$ and $w = 0.5$. In this case also a choice of $w = 0.5$ gives more weight than necessary to the smoothness thus resulting in a lower performance than the pure LA. This includes also the JDMIOA, which suggests that a Tikhonov prior should be considered in high noise regimes.

As a general observation w.r.t. the DAD algorithm, we can say that even though we compute the joint denoising in the mean online, i.e., the graph filter output is averaged for every collected sample, our approach gives a better performance than the offline alternative when the number of samples is limited. The initial transient behavior is due to the fact that the filters are initialized to zero. On the other hand, both approaches perform the same when the number of samples increase.

Solving the joint Tikhonov denoising in the mean leads to a reduction of the computational and communication complexity as compared to the disjoint approach. This is due to the fact that the joint approach does not require many iterations between the nodes to reach a reasonable performance. For the same reason, the joint denoising approach is more efficient also in terms of latency.

As a particular case, the pure Tikhonov denoising error can be seen for $t = 1$, when only one signal realization is available, in the DAD algorithm. With respect to this case, we can notice that taking multiple realizations into account can improve the performance up to one order of magnitue in only 10 iterations (Fig. 2(b), $w = 2$). To conclude, notice that the proposed approach even with a node variance of $\sigma_n^2 = 1$ and with 10 samples, can achieve an error of $10^{-2}$ for the signal of interest in only 10 iterations.

### C. Stochastic sparsification

We now continue with showing some simulation results when a desired operation is not carried out over the given deterministic graph $\mathcal{G}$, but on its sparsified version according to the results shown in Section V-B. For this purpose, we consider a graph signal characterized by a white unitary graph spectrum with reference to the graph $\mathcal{G}$. We derive the results for both the FIR [7] and ARMA [13] graph filters for $K = 1,\ 3,\ 5,\ 7$ designed as ideal low-pass filters with cut-off frequency equal to half the signal bandwidth. We use the same error measuring criterion as in (31) where now $\boldsymbol{z}^{(\mathrm{s})}$ will indicate the filter output of the sparsified filter.

Since choices of $\boldsymbol{L}$ with a low spectral norm are preferred in order to have a lower variance of the filter output and a larger stability region for the ARMA filters (and as a consequence higher approximation accuracy), we perform our results for two different scaled Laplacians, i.e., $\boldsymbol{L} = \boldsymbol{L}_{\mathrm{n}} - \boldsymbol{I}$ and $\boldsymbol{L} = \frac{1}{\lambda_{\max}}\boldsymbol{L}_{\mathrm{d}} - 0.5\boldsymbol{I}$ with $\lambda_{\max}$ the maximum eigenvalue of $\boldsymbol{L}_{\mathrm{d}}$ of the underlying graph $\mathcal{G}$. In case the latter Laplacian is used, the filter coefficients can be changed according to the results proposed in Section V-B. Meanwhile, for $\boldsymbol{L} = \boldsymbol{L}_{\mathrm{n}} - \boldsymbol{I}$ this cannot be done since we do not have a closed form expression of its expected value. In this instance, we will run the filters with the same filter coefficients as used in calculating the determinstic output, i.e., they will not change. As a consequence the zero mean error is not guaranteed in this case.

In Fig. 3, we show the mean error for different values of $p$ and for $\boldsymbol{L} = \boldsymbol{L}_{\mathrm{n}} - \boldsymbol{I}$. We can notice that the introduced bias in this case is very small and it reduces furthermore when the filter order is lower. Generally, if $p \geq 0.5$ we have that the mean error is smaller than $10^{-2}$.

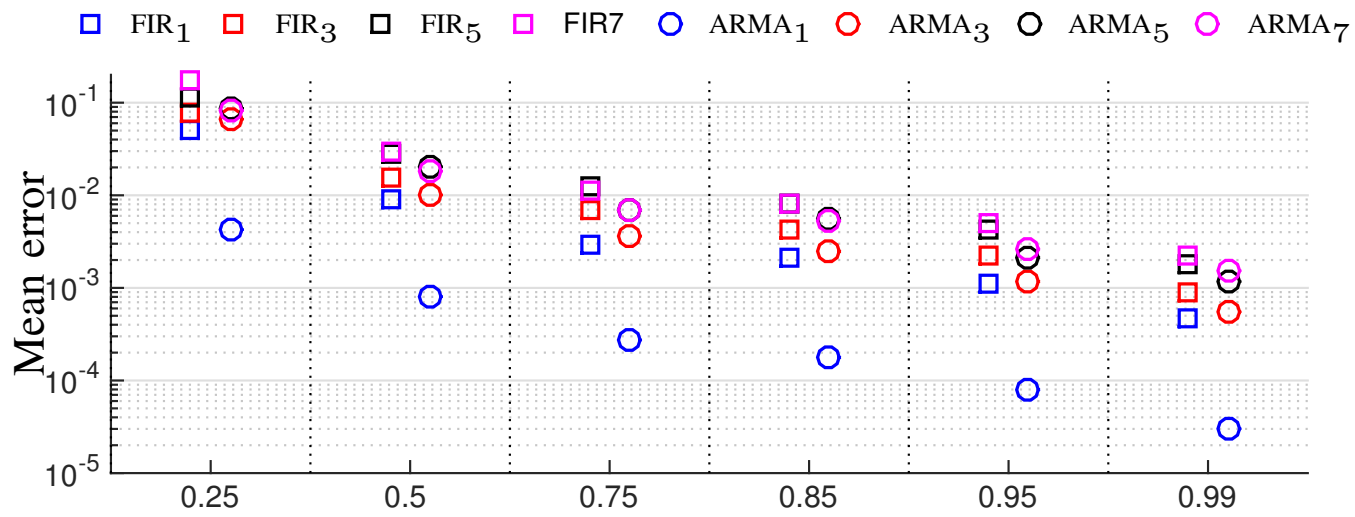

Fig. 3: Mean error over all nodes and realizations between sparsified and deterministic output for different values of $p$ when $\boldsymbol{L} = \boldsymbol{L}_{\mathrm{n}} - \boldsymbol{I}$.

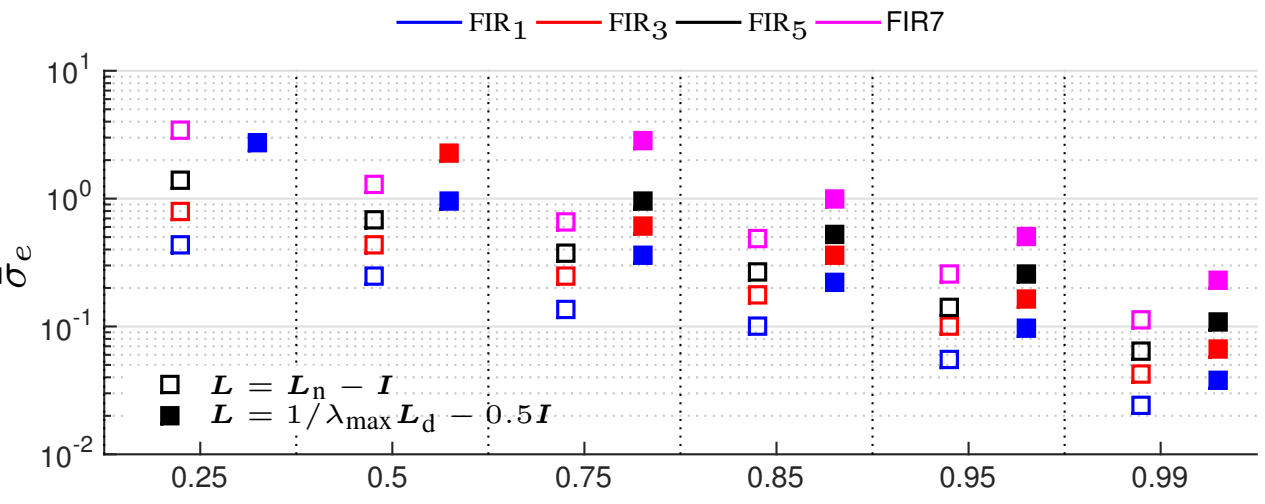

Fig. 4: Average empirical standard deviation among all nodes of the error between the sparsified FIR filter output and the original graph output for different values of $p$. The results are shown for two different Laplacians. For $\boldsymbol{L} = \boldsymbol{L}_{\mathrm{n}} - \boldsymbol{I}$ the filter coefficients are not changed.

Figs. 4 and 5 show the empirical standard deviation of the considered error for different values of $p$. As a common phenomenon for all filters, we can see that the standard deviation of the error reduces with higher values of $p$. Further, for all filters we can notice that a lower filter order yields a lower standard deviation. A notable result is observed for the ARMA$_1$ filter, where we can solve tasks like Tikhonov denoising, signal interpolation under smoothness assumptions, some Wiener based denoising and graph diffusion processes, with communication and computational costs reduced by 75% with little or no error. Indeed, for $p = 0.25$, the average standard deviation is $\bar{\sigma}_e \approx 0.035$ and a mean error smaller than 0.004. This result indicates that the signal output obtained in the sparsified graph, is closer to the same output obtained operating on underlying graph deterministically. For higher filter orders, we suggest $p$ should be around 0.75 to achieve a reasonable performance.

Similar to the results obtained in Fig. 1, the ARMA filters provide a lower $\bar{\sigma}_e$ than the FIR filters for the presented orders, while for higher values of $K$ the FIR filters seem to perform better. To conclude we can see that the choice of $\boldsymbol{L} = \boldsymbol{L}_{\mathrm{n}} - \boldsymbol{I}$ without changing the filter coefficients yields a good performance which suggests that the normalized Laplacian is a good choice to be robust to link fluctuations.

## VII. Conclusions

This work extends the analysis of graph filters to the stochastic setting. Based on the statistical knowledge of the graph and signal distribution, we characterize the first and second order moments of the output signal. We show that both the FIR and the ARMA graph filters behave in the mean as the same deterministic filter, having as input the mean signal, operating on a deterministic graph being the expected graph. We further show that when the graph signal is a stochastic process with a time-varying mean and covariance, the graph filters behave as a 2-dimensional filter and they capture the variations in the mean of the graph signal on the expected graph and time jointly. For both FIR and ARMA graph filters, we prove that the variance of the filter output is upper bounded and numerical results show that the empirical variance in different simulated scenarios is relatively small. Thus the signal realizations remain close to the expected value. The work is concluded by leveraging the stochasticity as a tool to improve denoising tasks and to perform stochastically sparsified graph filtering.

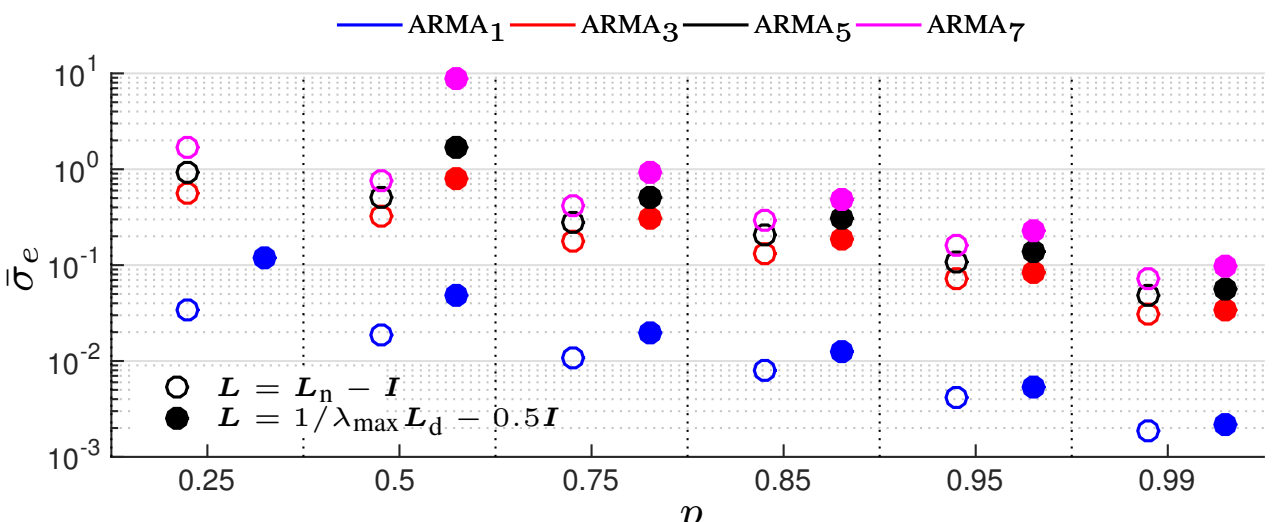

Fig. 5: Average empirical standard deviation among all nodes of the error between the sparsified ARMA filter output and the original graph output for different values of $p$. The results are shown for two different Laplacians. For $\boldsymbol{L} = \boldsymbol{L}_{\mathrm{n}} - \boldsymbol{I}$ the filter coefficients are not changed. Notice that for the ARMA$_1$ filter (useful for diffusion, interpolation and denoising) we can save up to 75% of communication and computational costs.

## Appendix

*Proof of Proposition 1*

By applying the expectation operator to (5) and considering that the realizations of the graph topology and graph signal are independent we have

$$\bar{\boldsymbol{z}}_{t+1} = \mathbb{E}\left[\sum_{k=0}^{K} \varphi_k \boldsymbol{\Phi}_{\boldsymbol{L}}(t, t-k+1)\boldsymbol{x}_{t-k+1}\right]$$
$$= \sum_{k=0}^{K} \varphi_k \left(\prod_{\tau=t}^{t-k+1} \mathbb{E}[\boldsymbol{L}_\tau]\right) \mathbb{E}[\boldsymbol{x}_{t-k+1}] = \sum_{k=0}^{K} \varphi_k \bar{\boldsymbol{L}}^k \bar{\boldsymbol{x}}, \quad (33)$$

where $\bar{\boldsymbol{z}}_{t+1}$ is the constant expected value of $\boldsymbol{z}_{t+1}$ after $K$ time steps. In (33) we substitute the back going product $\boldsymbol{\Phi}_{\boldsymbol{L}}(t, t-k+1)$ and apply the linearity of the expectation. □

*Proof of Theorem 1*

Recursion (7) can be differently written as

$$\boldsymbol{y}_{t+1} = (\boldsymbol{\Psi} \otimes \boldsymbol{L}_t)\boldsymbol{y}_t + \boldsymbol{\varphi} \otimes \boldsymbol{x}_t \quad (34a)$$
$$\boldsymbol{z}_{t+1} = (\boldsymbol{1}^\top \otimes \boldsymbol{I}_N)\boldsymbol{y}_{t+1}, \quad (34b)$$

with $\boldsymbol{y}_t = \left[\boldsymbol{y}_t^{(1)\top}, \boldsymbol{y}_t^{(2)\top}, \cdots, \boldsymbol{y}_t^{(K)\top}\right]^\top$ the $NK \times 1$ stacked state vector, $\boldsymbol{\Psi} = \mathrm{diag}(\psi^{(1)}, \psi^{(2)}, \cdots, \psi^{(K)})$ a diagonal $K \times$ $K$ coefficient matrix, $\boldsymbol{\varphi} = [\varphi^{(1)}, \varphi^{(2)}, \cdots, \varphi^{(K)}]^\top$ a $K \times 1$ coefficient vector, and $\mathbf{1}$ the $K \times 1$ one-vector. Then, in the following we consider that all the branches of the parallel filter bank are stable, i.e., $|\psi^{(k)}| = \|\boldsymbol{\Psi}\|_\infty < 1/\varrho$ imposed in the filter design phase. Then, by using the linearity of the expectation, and the independency among time between the realizations of the random variables we rewrite (34) as

$$\bar{\boldsymbol{y}}_{t+1} = \mathbb{E}[(\boldsymbol{\Psi} \otimes \boldsymbol{L}_t)]\, \bar{\boldsymbol{y}}_t + \mathbb{E}[\boldsymbol{\varphi} \otimes \boldsymbol{x}_t] \tag{35a}$$

$$\bar{\boldsymbol{z}}_{t+1} = (\mathbf{1}^\mathsf{T} \otimes \boldsymbol{I}_N)\bar{\boldsymbol{y}}_{t+1}, \tag{35b}$$

where $\bar{\boldsymbol{z}}_{t+1}$ and $\bar{\boldsymbol{y}}_{t+1}$ denote the expected values of $\boldsymbol{z}_{t+1}$ and $\boldsymbol{y}_{t+1}$, respectively. Then by using once again the independency of the graph realizations and graph signal together with the properties of the Kronecker product we rewrite (35) as

$$\bar{\boldsymbol{y}}_{t+1} = (\boldsymbol{\Psi} \otimes \bar{\boldsymbol{L}})\bar{\boldsymbol{y}}_t + \boldsymbol{\varphi} \otimes \bar{\boldsymbol{x}} \tag{36a}$$

$$\bar{\boldsymbol{z}}_{t+1} = (\mathbf{1}^\mathsf{T} \otimes \boldsymbol{I}_N)\bar{\boldsymbol{y}}_{t+1}, \tag{36b}$$

with $\bar{\boldsymbol{x}}$ being the expected values of $\boldsymbol{x}_t$. Then we expand (36) to express it as a function of the initial condition $\boldsymbol{y}_0$ and the inputs as

$$\begin{aligned}\bar{\boldsymbol{y}}_{t+1} &= \left(\boldsymbol{\Psi} \otimes \bar{\boldsymbol{L}}\right)^{t+1} \boldsymbol{y}_0 + \sum_{\tau=0}^{t} \left(\boldsymbol{\Psi} \otimes \bar{\boldsymbol{L}}\right)^{\tau} (\boldsymbol{\varphi} \otimes \bar{\boldsymbol{x}}) \\ &= \left(\boldsymbol{\Psi}^{t+1} \otimes \bar{\boldsymbol{L}}^{t+1}\right) \boldsymbol{y}_0 + \sum_{\tau=0}^{t} \left(\boldsymbol{\Psi}^{\tau}\boldsymbol{\varphi}\right) \otimes \left(\bar{\boldsymbol{L}}^{\tau}\bar{\boldsymbol{x}}\right),\end{aligned} \tag{37}$$

where we have used the Kronecker product property

$$(\boldsymbol{A} \otimes \boldsymbol{B})(\boldsymbol{C} \otimes \boldsymbol{D}) = (\boldsymbol{AC}) \otimes (\boldsymbol{BD}). \tag{38}$$

By using once again the stability condition for all the branches, we have that $\lim_{t\to\infty}\left\|\left(\boldsymbol{\Psi}^{t+1} \otimes \bar{\boldsymbol{L}}^{t+1}\right)\boldsymbol{y}_0\right\| = 0$. Hence, the limiting steady state of the expected value can be written as

$$\begin{aligned}\bar{\boldsymbol{z}} &= \lim_{t\to\infty} \sum_{\tau=0}^{t} (\mathbf{1}^\mathsf{T} \otimes \boldsymbol{I}_N)\left(\boldsymbol{\Psi}^{\tau}\boldsymbol{\varphi}\right) \otimes \left(\bar{\boldsymbol{L}}^{\tau}\bar{\boldsymbol{x}}\right) \\ &= \lim_{t\to\infty} \sum_{\tau=0}^{t} \left(\mathbf{1}^\mathsf{T}\boldsymbol{\Psi}^{\tau}\boldsymbol{\varphi}\right) \otimes \left(\bar{\boldsymbol{L}}^{\tau}\bar{\boldsymbol{x}}\right) \\ &= \lim_{t\to\infty} \sum_{\tau=0}^{t}\sum_{k=1}^{K} \varphi^{(k)} \left(\psi^{(k)}\bar{\boldsymbol{L}}\right)^{\tau} \bar{\boldsymbol{x}}\end{aligned} \tag{39}$$

where once again we made use of (38) and expressed the Kronecker product as the sum of $K$ terms. By leveraging once again the fact that all the branches of the filter bank are stable ($|\psi^{(k)}| < 1/\varrho$), we rewrite (39) as

$$\bar{\boldsymbol{z}} = \sum_{k=1}^{K} \varphi^{(k)} \left(\boldsymbol{I} - \psi^{(k)}\bar{\boldsymbol{L}}\right)^{-1} \bar{\boldsymbol{x}} \tag{40}$$

which proves the first part (8). Then from (37) we can see that the dependency on the initial state $\boldsymbol{y}_0$ decreases exponentially with $t$, thus we can say that (7) converges linearly to the limiting steady state value of the expected value (8). □

*Proof of Proposition 2*

By using the same arguments as in the proof of Proposition 1, one finds that the output of an $\text{FIR}_K$ is in expectation

$$\bar{\boldsymbol{z}}_{t+1} = \sum_{k=0}^{K} \varphi_k \bar{\boldsymbol{L}}^k \bar{\boldsymbol{x}}_{t-k+1}. \tag{41}$$

Then, by projecting the signal using the GFT into the subspace spanned by an eigenvector $\boldsymbol{\phi}_n$ of $\bar{\boldsymbol{L}}$ with associated eigenvalue $\lambda_n$ (41) is reduced to

$$\bar{z}_{t+1} = \sum_{k=0}^{K} \varphi_k \lambda_n^k \bar{x}_{t-k+1}, \tag{42}$$

where $\bar{x}_t = \boldsymbol{\phi}_n^\top \bar{\boldsymbol{x}}_t$ and $\bar{z}_{t+1} = \boldsymbol{\phi}_n^\top \bar{\boldsymbol{z}}_{t+1}$ are respectively the magnitude of the projections of the filter input and output on the chosen subspace. By taking the Z-transform and dividing both sides by $z^{t+1}$ the claim (10) follows. □

*Proof of Theorem 2*

By rewriting the parallel $\text{ARMA}_K$ in the form (34) and by following the same procedure as in the derivation of (36), we have that the expected ARMA output at time instant $t+1$ is

$$\begin{aligned}\bar{\boldsymbol{y}}_{t+1} &= \mathbb{E}[(\boldsymbol{\Psi} \otimes \boldsymbol{L}_t)\boldsymbol{y}_t + \boldsymbol{\varphi} \otimes \boldsymbol{x}_t] = (\boldsymbol{\Psi} \otimes \bar{\boldsymbol{L}})\bar{\boldsymbol{y}}_t + \boldsymbol{\varphi} \otimes \bar{\boldsymbol{x}}_t \\ \bar{\boldsymbol{z}}_{t+1} &= \mathbb{E}\big[(\mathbf{1}^\mathsf{T} \otimes \boldsymbol{I}_N)\boldsymbol{y}_{t+1}\big] = (\mathbf{1}^\mathsf{T} \otimes \boldsymbol{I}_N)\bar{\boldsymbol{y}}_{t+1}.\end{aligned} \tag{43}$$

In analogy to [13], (43) represents a deterministic 2-dimensional $\text{ARMA}_K$ filter with deterministic time-varying input signal $\bar{\boldsymbol{x}}_t$, filter memory $\bar{\boldsymbol{y}}_t$ and output $\bar{\boldsymbol{z}}_t$ operating over the deterministic time-invariant graph $\bar{\boldsymbol{L}}$. With these analogies (11) can be proven from Theorem 3 in [13].

*Proof of Proposition 3*

We start by computing the trace of the covariance matrix of the filter output at time instant $t+1$ as

$$tr(\boldsymbol{\Sigma}_{\boldsymbol{z}}[t+1]) = tr(\mathbb{E}\big[\boldsymbol{z}_{t+1}\boldsymbol{z}_{t+1}^\mathsf{H}\big]) - tr(\mathbb{E}[\boldsymbol{z}_{t+1}]\,\mathbb{E}[\boldsymbol{z}_{t+1}]^\mathsf{H}). \tag{44}$$

Using the linearity of the expectation, the first term on the right hand side of (44) can be expanded as

$$tr\left(\mathbb{E}\big[\boldsymbol{z}_{t+1}\boldsymbol{z}_{t+1}^\mathsf{H}\big]\right) = \sum_{k=0,l=0}^{K} \varphi_k\varphi_\ell T(k,\ell), \tag{45}$$

where

$$T(k,\ell) = tr\left(\mathbb{E}\big[\boldsymbol{\Phi}_{\boldsymbol{L}}(t,t-k+1)\boldsymbol{x}_{t-k+1}\boldsymbol{x}_{t-\ell+1}^\mathsf{H}\boldsymbol{\Phi}_{\boldsymbol{L}}(t,t-l+1)^\mathsf{H}\big]\right). \tag{46}$$

To proceed, note that by the commutativity of the trace with respect to the expectation and using the cyclic property of the trace, we can write

$$\begin{aligned}T(k,\ell) &= \mathbb{E}\big[tr\left((\boldsymbol{\Phi}_{\boldsymbol{L}}(t,t-l+1)^\mathsf{H}\boldsymbol{\Phi}_{\boldsymbol{L}}(t,t-k+1)\boldsymbol{x}_{t-k+1}\boldsymbol{x}_{t-\ell+1}^\mathsf{H}\right)\big] \\ &= tr\left(\mathbb{E}\big[\boldsymbol{\Phi}_{\boldsymbol{L}}(t,t-l+1)^\mathsf{H}\boldsymbol{\Phi}_{\boldsymbol{L}}(t,t-k+1)\big]\,\mathbb{E}\big[\boldsymbol{x}_{t-k+1}\boldsymbol{x}_{t-\ell+1}^\mathsf{H}\big]\right)\end{aligned} \tag{47}$$

with matrix

$$\mathbb{E}\big[\boldsymbol{x}_{t-k+1}\boldsymbol{x}_{t-\ell+1}^{\mathsf{H}}\big] = \begin{cases} \boldsymbol{\Sigma}_{\boldsymbol{x}} + \bar{\boldsymbol{x}}\bar{\boldsymbol{x}}^{\mathsf{H}}, & \text{if } k=\ell, \\ \bar{\boldsymbol{x}}\bar{\boldsymbol{x}}^{\mathsf{H}}, & \text{otherwise.} \end{cases} \quad (48)$$

being positive semi-definite. We can therefore use inequality

$$tr\,(\boldsymbol{A}\boldsymbol{B}) \leq \frac{\|\boldsymbol{A}+\boldsymbol{A}^{\mathsf{H}}\|}{2} tr(\boldsymbol{B}) \leq \|\boldsymbol{A}\| tr(\boldsymbol{B}) \quad (49)$$

valid for any square matrix $\boldsymbol{A}$ and positive semi-definite matrix $\boldsymbol{B}$ ($\boldsymbol{B} \succeq 0$) of appropriate dimensions [45], together with the triangle inequality and the fact that the realizations of the graph Laplacian are upper bounded as $\|\boldsymbol{L}_t\| \leq \varrho$ to bound $T(k,\ell)$

$$\begin{aligned} T(k,\ell) &\leq tr\left(\mathbb{E}\big[\boldsymbol{x}_{t-k+1}\boldsymbol{x}_{t-\ell+1}^{\mathsf{H}}\big]\right) \|\mathbb{E}\left[\Big(\prod_{\tau=t}^{t-\ell+1}\boldsymbol{L}_\tau\Big)\Big(\prod_{\tau=t}^{t-k+1}\boldsymbol{L}_\tau\Big)\right]\| \\ &\leq tr\left(\boldsymbol{\Sigma}_{\boldsymbol{x}} + \bar{\boldsymbol{x}}\bar{\boldsymbol{x}}^{\mathsf{H}}\right)\varrho^{k+\ell}, \end{aligned} \quad (50)$$

where in (50) we have also applied the Jensen's inequality for the spectral norm ($\|\mathbb{E}[\boldsymbol{A}]\| \leq \mathbb{E}[\|\boldsymbol{A}\|]$). We now can notice that the second term in the right-hand side of (44) is always positive. Thus it can be lower bounded as[4]

$$tr(\mathbb{E}[\boldsymbol{z}_{t+1}]\,\mathbb{E}\big[\boldsymbol{z}_{t+1}^{\mathsf{H}}\big] \geq 0. \quad (51)$$

Then, combining the results (50), (51) and (45) we can upper bound (44) as

$$tr(\boldsymbol{\Sigma}_{\boldsymbol{y}}[t+1]) \leq \sum_{k=0,l=0}^{K} \varphi_k\varphi_\ell\, tr\left(\boldsymbol{\Sigma}_{\boldsymbol{x}} + \bar{\boldsymbol{x}}\bar{\boldsymbol{x}}^{\mathsf{H}}\right)\varrho^{k+\ell}, \quad (52)$$

which then can be reformulated as (13) dividing both sides by $N$ and with simple algebra. □

*Proof of Theorem 3*

Let us set $\boldsymbol{C} = \boldsymbol{1}^{\mathsf{T}} \otimes \boldsymbol{I}_N$ for brevity. For the parallel ARMA$_K$ (34), we can express

$$\lim_{t\to\infty} tr(var(\boldsymbol{z}_{t+1})) = \lim_{t\to\infty} tr(var(\boldsymbol{C}\boldsymbol{y}_{t+1})) \quad (53)$$

Then, applying the definition and using the linearity of expectation, we have

$$\begin{aligned} var(\boldsymbol{C}\boldsymbol{y}_{t+1}) &= \boldsymbol{C}var(\boldsymbol{y}_{t+1})\boldsymbol{C}^{\mathsf{T}} \\ &= \boldsymbol{C}\mathbb{E}\big[\boldsymbol{y}_{t+1}\boldsymbol{y}_{t+1}^{\mathsf{H}}\big]\,\boldsymbol{C}^{\mathsf{T}} - \boldsymbol{C}\mathbb{E}[\boldsymbol{y}_{t+1}]\,\mathbb{E}[\boldsymbol{y}_{t+1}]^{\mathsf{H}}\,\boldsymbol{C}^{\mathsf{T}}. \end{aligned} \quad (54)$$

To proceed, we use inequality (49) and exploit the fact that $var(\boldsymbol{y}_{t+1}) \succeq 0$ is positive semidefinite matrix being the covariance matrix of $\boldsymbol{y}_{t+1}$. We can therefore apply (49) to (54) as

$$\begin{aligned} &\lim_{t\to\infty} tr(var(\boldsymbol{C}\boldsymbol{y}_{t+1})) = \lim_{t\to\infty} tr(\boldsymbol{C}^{\mathsf{T}}\boldsymbol{C}\,var(\boldsymbol{y}_{t+1})) \\ &\leq \lim_{t\to\infty} \|\boldsymbol{C}^{\mathsf{T}}\boldsymbol{C}\|\left(tr(\mathbb{E}\big[\boldsymbol{y}_{t+1}\boldsymbol{y}_{t+1}^{\mathsf{H}}\big]) - tr(\mathbb{E}[\boldsymbol{y}_{t+1}]\,\mathbb{E}[\boldsymbol{y}_{t+1}]^{\mathsf{H}})\right) \\ &= \lim_{t\to\infty} K\,tr(\mathbb{E}\big[\boldsymbol{y}_{t+1}\boldsymbol{y}_{t+1}^{\mathsf{H}}\big]) - K\,tr(\mathbb{E}[\boldsymbol{y}_{t+1}]\,\mathbb{E}[\boldsymbol{y}_{t+1}]^{\mathsf{H}}) \end{aligned} \quad (55)$$

where, in the last step, we made the substitution $\|\boldsymbol{C}^{\mathsf{T}}\boldsymbol{C}\| = \|\big(\boldsymbol{1}^{\mathsf{T}} \otimes \boldsymbol{I}_N\big)^{\mathsf{T}}\big(\boldsymbol{1}^{\mathsf{T}} \otimes \boldsymbol{I}_N\big)\| = K$ and we used the linearity of the expectation and trace: $\mathbb{E}[tr(\boldsymbol{A})] = tr(\mathbb{E}[\boldsymbol{A}])$. To ease notation, define $\boldsymbol{\Phi}(t',t) := \prod_{\gamma=t}^{t'} \boldsymbol{\Psi} \otimes \boldsymbol{L}_\gamma$ for $t' \geq t$, and $\boldsymbol{\Phi}(t',t) := \boldsymbol{I}$ if $t' < t$ and $\check{\boldsymbol{x}}_t = \boldsymbol{\varphi} \otimes \boldsymbol{x}_t$, after which the parallel ARMA$_K$ recursion (34a) becomes

$$\boldsymbol{y}_{t+1} = \boldsymbol{\Phi}(t,0)\boldsymbol{y}_0 + \sum_{\tau=0}^{t} \boldsymbol{\Phi}(t, t-\tau+1)\,\check{\boldsymbol{x}}_{t-\tau}. \quad (56)$$

After some algebraic manipulation and using the properties of the trace, we can write

$$\begin{aligned} \mathbb{E}\big[tr(\boldsymbol{y}_{t+1}\boldsymbol{y}_{t+1}^{\mathsf{H}}])) &= \mathbb{E}\Big[tr(\boldsymbol{\Phi}\,(t,0)^{\mathsf{H}}\,\boldsymbol{\Phi}\,(t,0)\,\boldsymbol{y}_0\boldsymbol{y}_0^{\mathsf{H}})\Big] \\ &+ \sum_{\tau=0}^{t}\mathbb{E}\Big[tr(\boldsymbol{\Phi}\,(t,t-\tau+1)^{\mathsf{H}}\,\boldsymbol{\Phi}\,(t,0)\,\boldsymbol{y}_0\check{\boldsymbol{x}}_{t-\tau}^{\mathsf{H}})\Big] \\ &+ \sum_{\tau=0}^{t}\mathbb{E}\Big[tr(\boldsymbol{\Phi}\,(t,0)^{\mathsf{H}}\,\boldsymbol{\Phi}\,(t,t-\tau+1)\,\check{\boldsymbol{x}}_{t-\tau}\boldsymbol{y}_0^{\mathsf{H}})\Big] \\ &+ \sum_{\tau_1,\tau_2=0}^{t}\mathbb{E}\Big[tr(\boldsymbol{\Phi}\,(t,t-\tau_2+1)^{\mathsf{H}}\,\boldsymbol{\Phi}\,(t,t-\tau_1+1)\,\check{\boldsymbol{x}}_{t-\tau_1}\check{\boldsymbol{x}}_{t-\tau_2}^{\mathsf{H}})\Big]. \end{aligned}$$

Due to the independence of signal and graph, as well as since $\mathbb{E}[\boldsymbol{y}_0] = 0$, the second and third terms above are equal to zero. Notice however that $\boldsymbol{y}_0\boldsymbol{y}_0^{\mathsf{H}} \succeq 0$ and $\check{\boldsymbol{x}}_{t-\tau_1}\check{\boldsymbol{x}}_{t-\tau_2}^{\mathsf{H}} \succeq 0$: $\boldsymbol{y}_0\boldsymbol{y}_0^{\mathsf{H}} \succeq 0$ is symmetric with one non-zero eigenvalue equal to $\|\boldsymbol{y}_0\|^2$, whereas $\mathbb{E}\big[\check{\boldsymbol{x}}_{t-\tau_1}\check{\boldsymbol{x}}_{t-\tau_2}^{\mathsf{H}}\big] \succeq 0$ is a Kronecker product of positive semi-definite matrices[5]

$$\begin{aligned} \mathbb{E}\big[\check{\boldsymbol{x}}_{t-\tau_1}\check{\boldsymbol{x}}_{t-\tau_2}^{\mathsf{H}}\big] &= \mathbb{E}\Big[(\boldsymbol{\varphi}\otimes\boldsymbol{x}_{t-\tau_1})\,(\boldsymbol{\varphi}\otimes\boldsymbol{x}_{t-\tau_2})^{\mathsf{H}}\Big] \\ &= \boldsymbol{\varphi}\boldsymbol{\varphi}^{\mathsf{H}} \otimes \mathbb{E}\big[\boldsymbol{x}_{t-\tau_1}\boldsymbol{x}_{t-\tau_2}^{\mathsf{H}}\big] \\ &= \begin{cases} \boldsymbol{\varphi}\boldsymbol{\varphi}^{\mathsf{H}} \otimes (\boldsymbol{\Sigma}_{\boldsymbol{x}} + \bar{\boldsymbol{x}}\bar{\boldsymbol{x}}^{\mathsf{H}}), & \text{if } \tau_1 = \tau_2, \\ \boldsymbol{\varphi}\boldsymbol{\varphi}^{\mathsf{H}} \otimes \bar{\boldsymbol{x}}\bar{\boldsymbol{x}}^{\mathsf{H}}, & \text{otherwise.} \end{cases} \end{aligned} \quad (57)$$

We can therefore use again inequality (49), as well as Jensen's inequality for the spectral norm ($\|\mathbb{E}[\boldsymbol{A}]\| \leq \mathbb{E}[\|\boldsymbol{A}\|]$), and the linearity of the expectation and the trace, to write

$$\begin{aligned} \mathbb{E}\big[tr(\boldsymbol{y}_{t+1}\boldsymbol{y}_{t+1}^{\mathsf{H}}])) &\leq \mathbb{E}\Big[\|\boldsymbol{\Phi}\,(t,0)^{\mathsf{H}}\|\|\boldsymbol{\Phi}\,(t,0)\|\Big]\; tr(\mathbb{E}\big[\boldsymbol{y}_0\boldsymbol{y}_0^{\mathsf{H}}\big]) \\ &+ \sum_{\tau_1,\tau_2=0}^{t}\mathbb{E}\Big[\|\boldsymbol{\Phi}\,(t,t-\tau_2+1)^{\mathsf{H}}\|\|\boldsymbol{\Phi}\,(t,t-\tau_1+1)\|\Big] \\ &\times tr(\mathbb{E}\big[\check{\boldsymbol{x}}_{t-\tau_1}\check{\boldsymbol{x}}_{t-\tau_2}^{\mathsf{H}}\big]) \end{aligned} \quad (58)$$

Next, we examine $\mathbb{E}[\|\boldsymbol{\Phi}(t_1,t_2)\|]$. Considering that, for all $p$-norms and for any matrix $\boldsymbol{A}$ and $\boldsymbol{B}$ $\|\boldsymbol{A}\otimes\boldsymbol{B}\| = \|\boldsymbol{A}\|\|\boldsymbol{B}\|$, the expected spectral norm of $\boldsymbol{\Phi}(t_1,t_2)$ is bounded by

$$\begin{aligned} \mathbb{E}[\|\boldsymbol{\Phi}(t_1,t_2)\|] &\leq \mathbb{E}\left[\prod_{\gamma=t_2}^{t_1}\|\boldsymbol{\Psi}\otimes\boldsymbol{L}_\gamma\|\right] = \mathbb{E}\left[\prod_{\gamma=t_2}^{t_1}\|\boldsymbol{\Psi}\|\|\boldsymbol{L}_\gamma\|\right] \\ &= \prod_{\gamma=t_2}^{t_1}\|\boldsymbol{\Psi}\|\,\mathbb{E}[\|\boldsymbol{L}_\gamma\|] = \prod_{\gamma=t_2}^{t_1}|\psi_{max}|\|\mathbb{E}[\boldsymbol{L}_\gamma]\| \\ &\leq (\varrho\,|\psi_{max}|)^{t_1-t_2+1}. \end{aligned} \quad (59)$$

[4]Notice that the goal is to present *a* bound. A tighter bound can be obtained using the results of Proposition 1 and writing $tr(\bar{\boldsymbol{z}}_{t+1}\bar{\boldsymbol{z}}_{t+1}^{\mathsf{H}}) = \sum_{k=0,\ell=0}^{K}\varphi_k\varphi_\ell\bar{\boldsymbol{x}}^{\mathsf{H}}\bar{\boldsymbol{L}}^{k+\ell}\bar{\boldsymbol{x}}$.

[5]If $\boldsymbol{A} \succeq 0$ and $\boldsymbol{B} \succeq 0$, then $\boldsymbol{A}\otimes\boldsymbol{B} \succeq 0$.

Therefore, in the limit, the first term of (58) vanishes.

$$\lim_{t\to\infty} \mathbb{E}\Big[\|\boldsymbol{\Phi}(t,0)^{\mathsf{H}}\|\|\boldsymbol{\Phi}(t,0)\|\Big] tr(\mathbb{E}[\boldsymbol{y}_0\boldsymbol{y}_0^{\mathsf{H}}]) \leq \lim_{t\to\infty} (\varrho\,|\psi_{max}|)^{2t+2} tr(\mathbb{E}[\boldsymbol{y}_0\boldsymbol{y}_0^{\mathsf{H}}]) = 0 \qquad (60)$$

The last step above is because the filter is stable $|\varrho\,\psi_{max}| \leq 1$.

Putting (58) and (59) together, while eliminating all terms that vanish, we obtain a bound for the first term of (55)

$$\begin{aligned}
&\lim_{t\to\infty} K\,\mathbb{E}\big[tr(\boldsymbol{y}_{t+1}\boldsymbol{y}_{t+1}^{\mathsf{H}}]))\\
&\leq \lim_{t\to\infty} K \sum_{\tau_1,\tau_2=0}^{t} (\varrho\,|\psi_{max}|)^{\tau_1+\tau_2}\, tr(\mathbb{E}\big[\check{\boldsymbol{x}}_{t-\tau_1}\check{\boldsymbol{x}}_{t-\tau_2}^{\mathsf{H}}\big])\\
&\leq \lim_{t\to\infty} K \sum_{\tau_1,\tau_2=0}^{t} (\varrho\,|\psi_{max}|)^{\tau_1+\tau_2}\, tr\left(\boldsymbol{\varphi}\boldsymbol{\varphi}^{\mathsf{H}} \otimes \left(\boldsymbol{\Sigma}_{\boldsymbol{x}} + \bar{\boldsymbol{x}}\bar{\boldsymbol{x}}^{\mathsf{H}}\right)\right)\\
&= K\, tr\left(\frac{\boldsymbol{\varphi}\boldsymbol{\varphi}^{\mathsf{H}} \otimes \left(\boldsymbol{\Sigma}_{\boldsymbol{x}} + \bar{\boldsymbol{x}}\bar{\boldsymbol{x}}^{\mathsf{H}}\right)}{(1-\varrho\,|\psi_{max}|)^2}\right), \qquad (61)
\end{aligned}$$

whereas for the remaining term from (54) we can see that it cannot be negative. Thus, we can lower bound[6] (notice that this term has to be subtracted) it as

$$\lim_{t\to\infty} K tr(\mathbb{E}[\boldsymbol{y}_{t+1}]\,\mathbb{E}[\boldsymbol{y}_{t+1}]^{\mathsf{H}}) \geq 0. \qquad (62)$$

Putting together (53), (61)-(62) and after some algebraic manipulation, we reach the bound

$$\lim_{t\to\infty} \overline{\mathrm{var}}[\boldsymbol{z}_{t+1}] \leq (K/N)\, tr\left(\frac{\boldsymbol{\varphi}\boldsymbol{\varphi}^{\mathsf{H}} \otimes \left(\boldsymbol{\Sigma}_{\boldsymbol{x}} + \bar{\boldsymbol{x}}\bar{\boldsymbol{x}}^{\mathsf{H}}\right)}{(1-\varrho\,|\psi_{max}|)^2}\right). \qquad (63)$$

We now use the property of the trace and Kronecker product, $tr(\boldsymbol{A}\otimes\boldsymbol{B}) = tr(\boldsymbol{A})tr(\boldsymbol{B})$ with $tr(\boldsymbol{\varphi}\boldsymbol{\varphi}^{\mathsf{H}}) = \|\boldsymbol{\varphi}\|^2$, since $\boldsymbol{\varphi}$ is a diagonal matrix. With the latter consideration (63) can be then reformulated as (14) with simple algebra. □

*Proof of Theorem 4*

The covariance matrix $\boldsymbol{\Sigma}_{\boldsymbol{z}}[t+1]$ of the filter output $\boldsymbol{z}_{t+1}$ in (16) can be expressed as

$$\begin{aligned}
\boldsymbol{\Sigma}_{\boldsymbol{z}}[t+1] &= \boldsymbol{C}\boldsymbol{\Sigma}_{\boldsymbol{y}}[t+1]\,\boldsymbol{C}^{\mathsf{T}}\\
&= \boldsymbol{C}\mathbb{E}\big[\boldsymbol{y}_{t+1}\boldsymbol{y}_{t+1}^{\mathsf{T}}\big]\,\boldsymbol{C}^{\mathsf{T}} - \boldsymbol{C}\bar{\boldsymbol{y}}_{t+1}\bar{\boldsymbol{y}}_{t+1}^{\mathsf{T}}\boldsymbol{C}^{\mathsf{T}},
\end{aligned} \qquad (64)$$

where in the above derivation we made use of the fact that $\boldsymbol{y}_t$ is independent from $\boldsymbol{x}_t$. Then, we can notice that only the autocorrelation matrix of the system memory $\boldsymbol{R}_{\boldsymbol{y}}[t+1] = \mathbb{E}\big[\boldsymbol{y}_{t+1}\boldsymbol{y}_{t+1}^{\mathsf{T}}\big]$ at time instant $t+1$ is unknown in (64). The term $\boldsymbol{C}\bar{\boldsymbol{y}}_{t+1}\bar{\boldsymbol{y}}_{t+1}^{\mathsf{T}}\boldsymbol{C}^{\mathsf{T}}$ can be recursively calculated using the statistical knowledge of time instant $t$. Substituting the expression of $\boldsymbol{y}_{t+1}$ in (17a) into $\boldsymbol{R}_{\boldsymbol{y}}[t+1]$ we have

$$\begin{aligned}
\boldsymbol{R}_{\boldsymbol{y}}[t+1] &= \bar{\boldsymbol{A}}\boldsymbol{R}_{\boldsymbol{y}}[t]\bar{\boldsymbol{A}}^{\mathsf{T}} + \bar{\boldsymbol{A}}\bar{\boldsymbol{y}}_t\bar{\boldsymbol{x}}_t^{\mathsf{T}}\boldsymbol{B}^{\mathsf{T}} + \boldsymbol{B}\bar{\boldsymbol{x}}_t\bar{\boldsymbol{y}}_t^{\mathsf{T}}\bar{\boldsymbol{A}}^{\mathsf{T}}\\
&+ \mathbb{E}_{\tilde{\boldsymbol{A}}}\left[\tilde{\boldsymbol{A}}_t\boldsymbol{R}_{\boldsymbol{y}}[t]\tilde{\boldsymbol{A}}_t^{\mathsf{T}}\right] + \boldsymbol{B}\left(\boldsymbol{\Sigma}_{\boldsymbol{x}}[t] + \bar{\boldsymbol{x}}_t\bar{\boldsymbol{x}}_t^{\mathsf{T}}\right)\boldsymbol{B}^{\mathsf{T}},
\end{aligned} \qquad (65)$$

where $\mathbb{E}\left[\tilde{\boldsymbol{A}}_t\boldsymbol{y}_t\boldsymbol{y}_t^{\mathsf{T}}\tilde{\boldsymbol{A}}_t^{\mathsf{T}}\right] = \mathbb{E}_{\tilde{\boldsymbol{A}}}\left[\tilde{\boldsymbol{A}}_t\mathbb{E}_{\boldsymbol{y}}\left[\boldsymbol{y}_t\boldsymbol{y}_t^{\mathsf{T}}\right]\tilde{\boldsymbol{A}}_t^{\mathsf{T}}\right]$ due to the independency between the graph realization at time instant $t+1$ and system memory at time instant $t$ and linearity of the expecation. Then, we can see that each entry of the matrix $\mathbb{E}_{\tilde{\boldsymbol{A}}}[\tilde{\boldsymbol{A}}_t\boldsymbol{R}_{\boldsymbol{y}}[t]\tilde{\boldsymbol{A}}_t^{\mathsf{T}}]$ is derived from the standard three matrix multiplication which can be expressed as (19). Let us now focus on the term $\boldsymbol{C}\bar{\boldsymbol{y}}_{t+1}\bar{\boldsymbol{y}}_{t+1}^{\mathsf{T}}\boldsymbol{C}^{\mathsf{T}}$. Using (17a) we can expand it as

$$\begin{aligned}
\boldsymbol{C}\bar{\boldsymbol{y}}_{t+1}\bar{\boldsymbol{y}}_{t+1}^{\mathsf{T}}\boldsymbol{C}^{\mathsf{T}} &= \boldsymbol{C}\bar{\boldsymbol{A}}\bar{\boldsymbol{y}}_t\bar{\boldsymbol{y}}_t^{\mathsf{T}}\bar{\boldsymbol{A}}^{\mathsf{T}}\boldsymbol{C}^{\mathsf{T}} + \boldsymbol{C}\bar{\boldsymbol{A}}\bar{\boldsymbol{y}}_t\bar{\boldsymbol{x}}_t^{\mathsf{T}}\boldsymbol{B}^{\mathsf{T}}\boldsymbol{C}^{\mathsf{T}}\\
&+ \boldsymbol{C}\boldsymbol{B}\bar{\boldsymbol{x}}_t\bar{\boldsymbol{y}}_t^{\mathsf{T}}\bar{\boldsymbol{A}}^{\mathsf{T}}\boldsymbol{C}^{\mathsf{T}} + \boldsymbol{C}\boldsymbol{B}\bar{\boldsymbol{x}}_t\bar{\boldsymbol{x}}_t^{\mathsf{T}}\boldsymbol{B}^{\mathsf{T}}\boldsymbol{C}^{\mathsf{T}}.
\end{aligned} \qquad (66)$$

Then, by substituting (65), (66) and $\mathbb{E}_{\tilde{\boldsymbol{A}}}[\tilde{\boldsymbol{A}}_t\boldsymbol{R}_{\boldsymbol{y}}[t]\tilde{\boldsymbol{A}}_t^{\mathsf{T}}]$ into (64) the claim (18) follows. □

[6] As for the FIR upper bound, also here one can exploit the structure of $\lim_{t\to\infty} K tr(\mathbb{E}[\boldsymbol{y}_{t+1}]\,\mathbb{E}[\boldsymbol{y}_{t+1}]^{\mathsf{H}})$ to achieve a tighter bound.